\documentclass[10pt]{iopart}

\usepackage[switch]{lineno}

\usepackage{graphicx}
\usepackage[utf8]{inputenc}
\usepackage{textcomp}

\usepackage{booktabs}
\usepackage{rotating}
\usepackage{multirow}
\usepackage{subfigure}

\usepackage[sort&compress,comma,square,numbers]{natbib}                                
\usepackage{natbib,url,hyperref}
\hypersetup{
     colorlinks   = true,
     citecolor    = blue,
     linkcolor    = blue,
     urlcolor     = blue
}

\usepackage{color}
\usepackage[usenames,dvipsnames,svgnames,table]{xcolor}

\begin{document}

\frenchspacing

\title{Growth mechanism for nanotips in high electric fields}

\author{Ville Jansson$^1$, Ekaterina Baibuz$^1$, Andreas Kyritsakis$^1$, Simon Vigonski$^{1,2}$, Vahur Zadin$^2$, Stefan Parviainen$^1$, Alvo Aabloo$^2$, and Flyura Djurabekova$^1$}

\address{$^1$ Helsinki Institute of Physics and Department of Physics, P.O. Box 43\\ (Pehr Kalms gata 2), FI-00014 University of Helsinki, Finland}

\address{$^2$ Institute of Technology, University of Tartu, Nooruse 1, 50411 Tartu, Estonia}

\ead{ville.b.c.jansson@gmail.com}

\begin{abstract}
In this work we show using atomistic simulations that the biased diffusion in high electric field gradients creates a mechanism whereby nanotips may start growing from small surface asperities. It has long been known that atoms on a metallic surface have biased diffusion if electric fields are applied and that microscopic tips may be sharpened using fields, but the exact mechanisms have not been well understood. Our Kinetic Monte Carlo simulation model uses a recently developed theory for how the migration barriers are affected by the presence of an electric field. All parameters of the model are physically motivated and no fitting parameters are used. The model has been validated by reproducing characteristic faceting patterns of tungsten surfaces that have in previous experiments been observed to only appear in the presence of strong electric fields. The growth effect is found to be enhanced by increasing fields and temperatures.

\vspace{1cm}
\noindent{\it Keywords\/}: Kinetic Monte Carlo, surface diffusion, electric fields, tungsten, nanotips, vacuum arcs

\end{abstract}

\maketitle
\ioptwocol	

\section{Introduction}

Electric fields may be used as an effective and precise way to manipulate surfaces on the atomic scale \cite{stroscio1991atomic,tsong2005mechanisms}. 
An applied electric field may affect the surface in many ways, such as heating it through resistive Joule heating \cite{parviainen2011electronic}, creating currents that drive the diffusion of vacancies or surface atoms due to an electron wind force \cite{stroscio1991atomic}, evaporation of atoms from the surface \cite{kyritsakis2018thermal}, or generally roughen the surface by disordering the atomic surface layers \cite{de2018electric,de2019electric}.
A powerful means of manipulation is to make surface atoms (adatoms) migrate with a biased random walk, where the bias appear when an electric field is applied.
How the single surface atom migration becomes biased in electric fields has only recently been explained theoretically \cite{kyritsakis2019atomistic}, but how larger atomic surface structures are affected by fields has not been fully explored. 
It has been experimentally observed that nanostructures, such as atom islands, surface nanotips or mounds, may appear in electric fields \cite{whitman1991manipulation,mendez1996diffusion,mayer1999electric,dulot2000stm}, but the precise processes and mechanisms have not been studied in detail.  
It is also known that the application of a high electric in combination with high temperature can be used to sharpen the metal tips used in various applications such as electron microscopy \cite{bettler1960activation, dyke1960electrical, tsong2005mechanisms}.

Of particular interest is the question whether high electric fields may create sharp nanotips.
Nanotips are nanofeatures that are of interest because of their optical, mechanical and electrical properties and can be used in many applications, such as electron sources \cite{zhou2005growth,bormann2015ultrafast}, bio-sensors \cite{cherevko2009gold,kabashin2009plasmonic,caldwell2011plasmonic}, plasmonic trapping for the manipulation of nanoscale objects \cite{wang2012plasmonic}, and biocompatible electrodes \cite{bruggemann2011nanostructured}. 
Importantly, spontaneous growth of nanotips in high electric fields is also hypothesised to reduce the performance of linear particle accelerators by triggering electric discharges (arcs or breakdowns) in the vacuum of e.g. accelerating structures \cite{navitski2013field,nagaoka2001field,engelberg2018stochastic,clic2018compact}.
Nanotips with high enough aspect ratio, sufficiently enhance the field at their apex to cause field emission of electrons and neutrals that in turn may form a vacuum arc plasma \cite{kyritsakis2018thermal,timko2014from}.  
The hypothesis that such nanotips may grow has, however, not yet been proven.
The phenomenon is hard to study experimentally as only craters are left after the vacuum arc events and any evidence of possible nanotips can be assumed to be destroyed \cite{wuensch2013advances}.

The single atom diffusion in electric fields has been studied extensively both experimentally and theoretically, especially by T\;T\;Tsong et al., who showed experimentally in Field-Ion Microscopy (FIM) studies that the adatom migration is biased along the field gradient towards stronger fields \cite{tsong1975direct,wang1982field}. 
Tsong et al. proposed that the migration energy barrier of a surface atom in a field is dependent of its atomic dipole moment and polarizability \cite{tsong1975direct}. 
In a recent paper \cite{kyritsakis2019atomistic}, we refined this theory and developed a method for calculating the migration energy barriers in electric fields with Density Functional Theory (DFT).
By comparing with DFT calculations of the migration barriers of a W adatom on a W\{110\} surface, we were able to show that it is not enough to only consider the adatom's dipole moment and polarizability, but that it is also necessary to take into account the redistribution of the charge density of the whole surrounding system in order to calculate the correct barrier.

Knowing the correct theory for how electric field gradients create a bias in the atom migration processes enables the possibility of implementing this into a Kinetic Monte Carlo (KMC) model which would possibly allow the study of the surface evolution in electric fields on a larger scale. 
For just the thermal evolution of atomic surfaces, there already exist many KMC models \cite{hakkinen1993roughening,wang2001kinetic,lam2002competing,zhang2004kinetic,kara2009off,nandipati2012off} and in particular the open-source KMC code Kimocs \cite{jansson2016long, kimocs} has been used to study many different nanosystems and -structures, such as nanotips \cite{jansson2016long}, nanowires \cite{vigonski2018au} and nanoclusters \cite{zhao2016formation}. 
However, there are to date no published KMC models for surface diffusion processes that incorporates the effects of electric fields.
Without fields, Cu nanotips have been found in KMC simulations to flatten down due to surface diffusion \cite{jansson2016long}. 
In Molecular Dynamics simulations, which are limited to the nanoseconds timescale, both adatom migration \cite{vurpillot2018simulation} and nanotips in fields have been studied \cite{djurabekova2011atomistic, parviainen2011electronic,parviainen2011molecular,veske2016electrodynamics,kyritsakis2018thermal}. 
For very large nanotips ($\sim$90 nm) in high fields, a growth process have been observed where the tip is dragged upwards and stretched by the field \cite{kyritsakis2018thermal}. 
However, this process does not explain how a nanotip would be created in the first place. 
For initiating the formation of a tip, diffusion is a more likely mechanism, as W tip shapening has been observed experimentally \cite{bettler1960activation, dyke1960electrical, tsong2005mechanisms}.
It is worth noting that nanotips have also been observed to grow without any field present, but only in systems with interfaces between two materials: copper and carbon \cite{wang2004novel}, and iron and sapphire \cite{amram2015capillary}.


In this work, we will show using KMC simulations that the application of a high electric field on a tungsten surface may cause diffusion-driven growth of nanotips. 
For this purpose, we have developed a KMC model where the field effects on the migration energy barriers are based on recent theoretical works. 
In order to validate our model, we will also show that our model is able to reproduce previous experimental results where a tungsten tip showed characteristic facet patterns at different applied electric field strengths. 

The paper is structured as follows: The theoretical basis of the model is summarized in section \ref{sec:theory} and its KMC implementation is described in section \ref{sec:methods}. 
In section \ref{sec:fujita}, we will validate our model by reproducing the particular faceting of a W tip that has been experimentally observed to occur in high electric fields.
In section \ref{sec:nanotip}, we will investigate the mechanism for how nanotips may grow in electric fields and we will show that an W hemispherical asperity in a strong enough electric field and high enough temperature may start to grow and form a nanotip.
In section \ref{sec:sensitivity} we will show the robustness of the results by studying the sensitivity of the calculated values of the model's central parameters.
Finally, we discuss the findings in section \ref{sec:discussion} and summarize our conclusions in section \ref{sec:conclusions}.

\section{Theory}\label{sec:theory}

The probability rate for a surface atom to make a transition to another position is given by the Arrhenius equation 
\begin{equation}\label{eq:arrhenius}
\Gamma = \nu\exp \left(\frac{-E_m}{k_B T}\right),
\end{equation}
where $T$ is the temperature of the system and $k_B$ is the Boltzmann constant. 
The attempt frequency $\nu$ and the migration energy (or activation energy) $E_m$ will vary depending on the type of transition, but it is common to use the same attempt frequency for any transition, especially if only first-nearest neighbour jumps are used (see e.g. \cite{soisson2007cu,vincent2008precipitation,castin2011modeling,jansson2016long,baibuz2018migration,vigonski2018au}).
We use the same $\nu$ also for second-nearest jumps and exchange processes, as discussed in \cite{jansson2020tungsten}. 
The migration energies $E_m$, which we for this work take from \cite{jansson2020tungsten}, will depend on the local environment of the transitions; 
i.e. how many atoms are in the vicinity of the transition and (to a lesser degree) how these neighbour atoms are positioned \cite{baibuz2018migration}. 
For surface atomic processes, it has been found that only considering the number of first- and second-nearest neighbour atoms is enough to characterise the migration energy barriers that are needed to construct KMC simulation models that are precise enough to reproduce Molecular Dynamics (MD) simulations \cite{jansson2016long, jansson2020tungsten} or experimental results \cite{vigonski2018au}. 

If an external electric field is applied on the surface, this will also affect the migration energy barrier $E_m$ of the surface atom, which will depend on the field strength, the field direction, the field gradient along the atom jump, the type of atom that is jumping and which elements it is surrounded by. This effect was studied theoretically and described in detail in \cite{kyritsakis2019atomistic}, so we will in the following only summarize the main findings. 

Applying an electric field on a metallic surface will rearrange the charge distribution $\rho(\vec r)$ of the atomic system. 
In an applied field, the $\rho(\vec r)$ will also change if the atomic configuration  changes, such as in the case of a surface atom making a transition to a neighbour lattice point. This change of $\rho(\vec r)$ will also change the system dipole moment, $\vec p = (p_x,p_y,p_z) = \int \rho \vec r dV$, which can be calculated for small atomic systems with DFT, where the system would include an atomic substrate with a free surface, on top of which the jumping adatom is placed \cite{kyritsakis2019atomistic}. The system needs to be wide enough for the local field enhancement of the adatom not to affect the surface fields at the borders of the system and the substrate needs to be thick enough for the field not to penetrate it. We assume the field to be uniform or having a uniform field gradient at the upper vacuum boarder, above the substrate. For an electric field $F = |\vec F|$ in the z direction, perpendicular to the surface, the energy of the system will be given by \cite{kyritsakis2019atomistic}
\begin{equation}\label{eq:K1}
 E(F) = E(0) - \int_0^F p_z(F')dF',
\end{equation}
where $E(0)$ is the energy without any external field. In \cite{kyritsakis2019atomistic}, it is shown that $p_z$ may for small fields be represented as a Taylor expansion
\begin{equation}
 p_z(F) = \mathcal{M} + \mathcal{A} F + O(F^2),
\end{equation}
where $\mathcal{M}$ is the permanent dipole moment and $\mathcal{A}$ is the polarizability of the system. The energy of the system, equation \ref{eq:K1}, may now be written as \cite{kyritsakis2019atomistic}
\begin{equation}
 E(F) = E(0) -\mathcal{M} F -\frac{1}{2}\mathcal{A} F^2 + O(F^3).
\end{equation}
The parameters $\mathcal{M}$ and $\mathcal{A}$ are calculable with DFT and will change when the atomic configuration of the system changes.
The migration energy barrier $E_m(F)$ for a surface atom transition jump in an applied electric field $F$ is given by \cite{kyritsakis2019atomistic}
\begin{equation}\label{eq:barrier_uniform_field}
 E_m(F) = E_s - E_l = E_m(0) - \mathcal{M}_\textrm{sl} F - \frac{1}{2} \mathcal{A}_\textrm{sl} F^2,
\end{equation}
where $E(0)$ is the barrier without any field, $\mathcal{M}_\textrm{sl} \equiv \mathcal{M}_s - \mathcal{M}_l$, and $\mathcal{A}_\textrm{sl} \equiv \mathcal{A}_s - \mathcal{A}_l$; 
where the subscript $s$ refer to the system with the transition atom in the saddle point position and the subscript $l$ refer to the same system, but with the transition atom in the (initial) lattice position. 
If the applied field is not homogeneous, but has a gradient $\gamma \equiv dF/dx$ along the transition direction, which is so small that the difference in the applied field is negligible within a cut-off radius, $R_c \sim$ 1--2 lattice constants, i.e. $\gamma R_c \ll F_l$, the migration energy barrier can be written as \cite{kyritsakis2019atomistic}
\begin{equation}\label{eq:barrier_gradient}
 E_m \approx E_m(0) - \mathcal{M}_\textrm{sl} F_l - \frac{\mathcal{A}_\textrm{sl}}{2} F_l^2 - \mathcal{M}_\textrm{sr} \Delta F - \mathcal{A}_\textrm{sr} F_l \Delta F,
\end{equation}
where $\Delta F \equiv F(x_s) - F(x_l) = \gamma (x_s - x_l)$. 
Here $\mathcal{M}_\textrm{sr} \equiv \mathcal{M}_s -\mathcal{M}_r$ and $\mathcal{A}_\textrm{sr} \equiv \mathcal{A}_s - \mathcal{A}_r$, where the subscript $r$ refer to the system without the transition atom, the reference system.
Equations \ref{eq:barrier_uniform_field} and \ref{eq:barrier_gradient} were confirmed for a W jump on a W\{110\} surface using DFT calculations in \cite{kyritsakis2019atomistic}. 
The $\mathcal{M}$ and $\mathcal{A}$ parameters were calculated for the first nearest neighbour self-diffusion jump on the W\{110\} and the values are reproduced in table \ref{table:field_parameters}.
\begin{table}
  \centering
   \caption{The electronic parameters for equation \ref{eq:barrier_gradient}, as calculated for the first-nearest neighbour jump of a W adatom on a W\{110\} surface with DFT \cite{kyritsakis2019atomistic}.}
   \label{table:field_parameters}
   \begin{tabular*}{\columnwidth}{@{\extracolsep{\fill}} l l r}
   \toprule
   \midrule
   $\mathcal{M}_\textrm{sl}$ & [e m]			& $-3.19\cdot10^{-12}$	\\
   $\mathcal{A}_\textrm{sl}$ & [e m$^2$ V$^{-1}$]	& $3.1\cdot10^{-22}$	\\
   $\mathcal{M}_\textrm{sr}$ & [e m]			& $2.735\cdot10^{-11}$ 	\\
   $\mathcal{A}_\textrm{sr}$ & [e m$^2$ V$^{-1}$]	& $2.6\cdot10^{-21}$	\\
   \bottomrule
   \end{tabular*}
\end{table}

\section{Methods}\label{sec:methods}

In this work, we used the open-source KMC code Kimocs \cite{jansson2016long} together with the field-solver from the code {\sc Helmod} \cite{djurabekova2011atomistic}. 
Kimocs was specially developed to study atom diffusion processes on metallic surfaces. 
It uses a rigid lattice for the atoms, that are either face-centred-cubic (fcc) or body-centred-cubic (bcc).
In this paper we use a bcc lattice and a W parameterization, described in \cite{jansson2020tungsten}, which includes first- and second-nearest jumps. 
The parameterization also includes a third-nearest neighbour diagonal exchange process, which has been found important for the \{100\} surface \cite{olewicz2014coexistence,jansson2020tungsten}.
The implementation in Kimocs treats exchange processes as normal jump processes, but uses the NEB-calculated migration energy barriers of the actual exchange process.
Since only atoms of the same element (here W) are used and the atoms thus are indistinguishable, the omission of the actual exchange will not change the evolution nor the dynamics of the system. 
The parameterization and its validation is described in detail in \cite{jansson2020tungsten}.
The attempt frequency $\nu = 4.3 \cdot 10^{14}$\;s$^{-1}$ was fitted to MD simulations in \cite{jansson2020tungsten} and is used for all processes.

The probability for any atom transition event is calculated using equation \ref{eq:arrhenius}. 
The migration energy barriers in the absence of any electric field $E_m(0)$ are different for all atom transition processes and need to be tabulated prior to the KMC simulations.
The individual atom transitions are described by counting the number of first and second nearest neighbours of the initial position ($a$ and $b$) as well as the final position ($c$ and $d$). 
For every process of the same transition distance with the same neighbour configuration $(a,b,c,d)$, the same pre-calculated migration energy $E_m(a,b,c,d)$ is used. 
We only consider pure metals in our simulations; vacancies are the only possible defects. 

Atoms or clusters of atoms are not allowed to be detached from the surface, since these would not be taken correctly into account in the field calculations (see below). 
Nor do we explicitly account for evaporation, but atoms may become isolated if the surrounding atoms are diffusing away. 
Such isolated atoms, with no nearest neighbours and one or less second-nearest neighbour atoms, will be removed from the system in order to avoid simulation artefacts.
We will however still refer to these removed atoms as ``evaporated'' for simplicity.

\subsection{The field solver}\label{sec:field_solver}
\begin{figure}
 \includegraphics[width=\columnwidth]{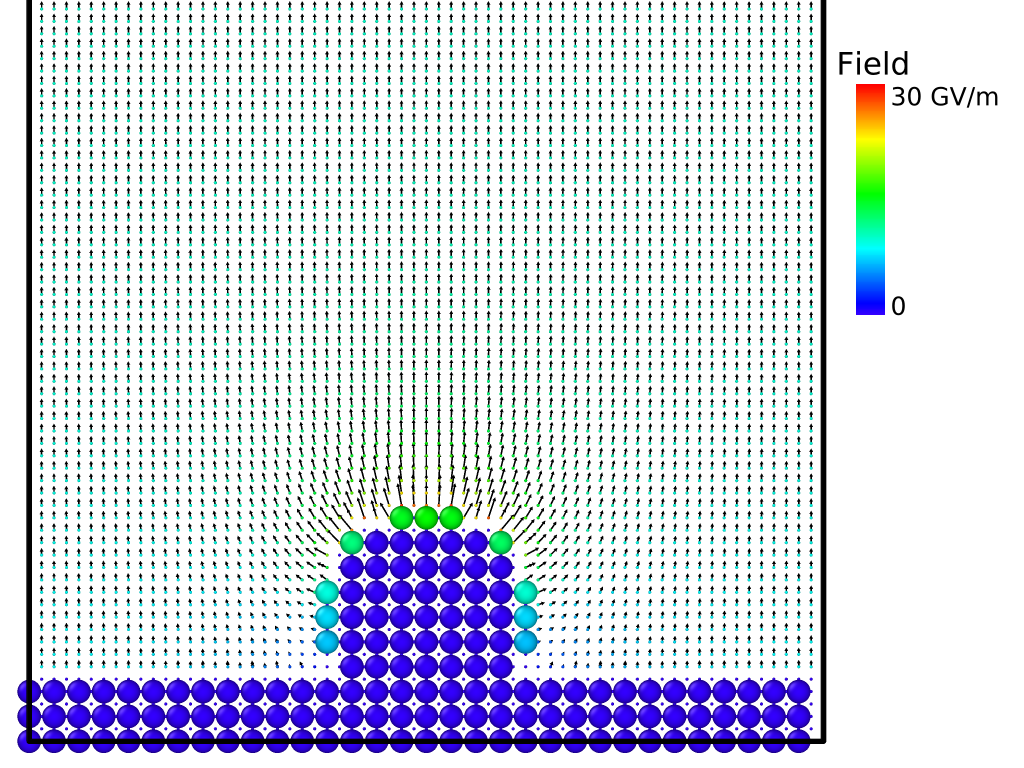}
 \caption{Cross-section of a system with a spherical asperity on a flat atomic surface. Atoms are shown as large spheres, whereas the dots and arrows are grid points in the vacuum, showing the field strengths and directions calculated by the field solver.}
 \label{fig:field_system}
\end{figure}

The field is calculated using the field model from the hybrid Molecular Dynamics-electrodynamics code, now known as {\sc Helmod}, which is thoroughly described in \cite{djurabekova2011atomistic}. 
In the {\sc Helmod} model, the electric field above an arbitrarily rough, but still continuous, metallic surface (see figure \ref{fig:field_system}) is calculated by solving Laplace's equation
\begin{equation}\label{eq:laplace}
 \nabla^2 \Phi = 0,
\end{equation}
where $\Phi$ is the electrostatic potential \cite{jackson1975classical}. 
Using Laplace's equation instead of the more general Poisson's equation, which takes into account space charges, is sufficient as we in this model focus on surface processes and do not consider atoms or ions detached from the surface. 
The field is not allowed to penetrate the surface more than 0.2 nm (about one atomic layer), as has been estimated for metal surfaces \cite{jackson1975classical}. 
Equation \ref{eq:laplace} is solved using the finite-difference method on a fine 3D grid with Gauss-Seidel iterations \cite{press1992numerical}.
The field grid is simple-cubic and aligned so that all (rigid) atom positions will coincide with the field grid points. 
For the cubic bcc unit cell with 2 atoms, there will be $2\times2\times2$ field grid points in the unit cell, of which 2 coincide with the atomic positions and 6 grid points are free.

At the lower boundary at the metal surface, a Dirichlet type condition, i.e. $\Phi = 0$, is used.
At the upper boundary of the vacuum, within which space the field is calculated, a Neumann type condition is used, i.e. $-\nabla \Phi = \pm F \hat{z}$, where $F$ is the external field. 
For the anode, we define the positive field $F > 0$ to be directed upwards (in the $+\hat z$ direction) and in the cathode case, the negative field $F < 0$ is directed downwards, towards the surface (in the $-\hat z$ direction).

\subsection{Implementation of the field effect into the KMC model}\label{sec:implementation}

In the KMC simulations, all possible atom transition jumps are assigned a probability rate according to the Arrhenius formula \ref{eq:arrhenius}, where the migration energy barrier $E_m$ is defined according to the field-dependent formula \ref{eq:barrier_gradient}. 
The field-independent barriers $E_m(0)$ are pre-calculated and tabulated (see \cite{jansson2020tungsten}). 

\begin{figure}
 \includegraphics[width=\columnwidth]{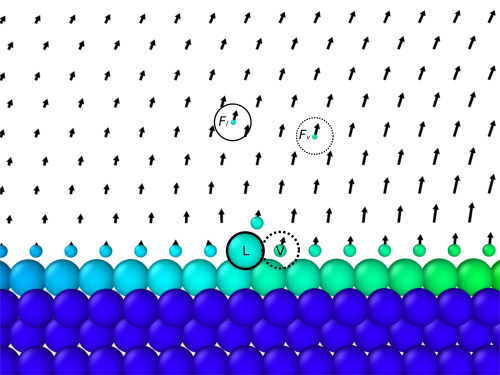}
 \caption{Cross-section of a surface under a field gradient. 
 Atoms are depicted as large spheres and surface vacancies as small spheres.
 Shown are also the field vectors, calculated by the field solver for the of the simple-cubic field lattice points.
 The surface atom $L$, marked by a circle, at a lattice position, will be assigned the field magnitude $F_l$ of a field point at a distance $h=1$\;nm above the surface, marked by a thin circle.
 Similarly, the surface vacancy $V$, marked with a dotted circle, is assigned the field magnitude $F_v$ from the position marked with a thin dotted circle, 1\;nm ($h$) along the field lines above the vacancy.
 The gradient for atom $L$ to jump to $V$ is given by $(F_v - F_l)/d$, where $d$ is the distance between the field points of $F_v$ and $F_l$.
 The colours of the atoms and vacancies corresponds to the assigned field magnitudes, with dark blue for the zero fields of the bulk atoms.}
 \label{fig:field}
\end{figure}

The surface field magnitude $F_l$ in equation \ref{eq:barrier_gradient}, which is the field assumed to be interacting with a surface atom at a lattice point position, is in Kimocs defined as the field value calculated by the field solver at a distance $h = 1$\;nm along the field lines, roughly perpendicularly, above the surface atom (figure \ref{fig:field}). 
This way the field value is only minimally affected by the local field enhancement of the atom itself, in accordance with the theoretical definition of $F_l$, as described in section \ref{sec:theory}. 
The direction of the field vector at the surface atom will be calculated by taking the average of the non-zero field vectors of the 26 surrounding field lattice points, as the field is not calculated by the field-solver for the points coinciding with atoms.
The field lattice points are in a simple-cubic field lattice and will have a spacing that depends on the atomic lattice; for a bcc lattice with a $\langle 100\rangle$ $z$ orientation, as in this work, the field lattice will have a lattice parameter $a_0/2$, where $a_0$ is the lattice parameter of the atomic bcc lattice. 
This way, all surface atoms will be assigned a field value $F_l$, which will be positive for an anode field and negative for a cathode field.

Surface vacancies, i.e. an empty atomic lattice positions with at least one first-nearest neighbour atom, will be assigned field values $F_v$ the same way, except that the direction of the field vector is taken directly from the field solver at the field grid points coinciding with the vacancies; i.e. no average is needed as for the atoms.
The magnitude of the fields are still taken at the distance $h$ above the surface vacancies.

For every surface atom, the field gradient $\gamma$ needs to be calculated in the direction of every possible atom jump.
In Kimocs, this is done as
\begin{equation}\label{eq:gradient}
\gamma = \frac{F_v - F_l}{d_f},
\end{equation}
where $F_l$ is the field of the jumping atom, $F_v$ is the field of the vacancy at the target lattice position, and $d_f$ is the distance between the two field lattice points at a distance $h$ above the surface where the magnitudes of $F_l$ and $F_s$ were actually calculated by the field solver (figure \ref{fig:field}).

The difference in field between the saddle point and the initial lattice point $\Delta F \equiv F_s - F_l$, as defined in equation \ref{eq:barrier_gradient}, is calculated as
\begin{equation}\label{eq:Delta_F}
 \Delta F = \gamma d,
\end{equation}
where $d$ is the distance between the saddle point and the initial position of the jumping atom.
In this work, we always assume the saddle point to be exactly half-way of any atom jump or exchange process.
The $\Delta F$ values have to be recalculated for every possible atom jump in the system every time the field has been recalculated by the field-solver, which happens after every KMC step. 
Calculating the field is the most costly part of the simulation and uses $\sim$99 \% of the CPU time.

To estimate the errors of $F_l$ and $\gamma$ (and by extension $\Delta F$), consider the simple system with an adatom on a perfectly flat surface under a field gradient, shown in figure \ref{fig:field}. Removing the adatom at $L$ and comparing the field strengths at positions $L$ and $V$ with the $F_l$ and $F_v$ values, respectively, gives a difference of 3.4 \%.
Comparing the values at the points of $F_l$ and $F_v$ with and without the adatom ($L$) present, gives a difference of less than 1 \%.
Comparing the gradients $\gamma$ with and without the adatom ($L$) present, gives a difference of 13 \%.
Since removing atoms individually in order to precisely calculate the field without their own field-enhancement would require too many calls to the computationally expensive field solver, these errors can be considered quite acceptable in comparison.

The $\mathcal{M}$ and $\mathcal{A}$ parameters in equation \ref{eq:barrier_gradient} were calculated for a first-nearest jump of a W atom on a W\{110\} in \cite{kyritsakis2019atomistic} and are reproduced in table \ref{table:field_parameters}. 
In this work these values are used for any atom jump or exchange process, as a first approximation, which is still good enough to enable the reproduction of experimental results, as will be shown in section \ref{sec:fujita}.

\section{Results}\label{sec:results}

\subsection{Model validation by comparison with experiments: faceting in fields}\label{sec:fujita}


S\;Fuijita and H\;Shimoyama have in \cite{fujita2007mechanism} experimentally demonstrated that a W tip will exhibit distinctly different energy equilibrium faceting patterns at different levels of applied electric fields.
In \cite{jansson2020tungsten} it was shown that our no-field W model correctly predicts the energy minimum shapes of W clusters, as given by MD simulations and Wulff constructions.
In this section we will show that including the field-effect into the model gives the same kind of energy equilibrium faceting patterns of a W nanotip at different electric fields as was observed by S\;Fujita and H\;Shimoyama.
We will simulate a smaller nanotip than was used in the experiment due to computational restrictions, but reproducing the same faceting stages at different levels of applied electric fields would still be a strong validation of our field model.

In S\;Fuijita and H\;Shimoyama's experiment \cite{fujita2007mechanism}, a W tip with a rounded apex with a curvature of $\sim$206\;nm was subjected to a sequence of anode fields of different strengths at 30 s periods in an ultra high vacuum.
The tip was kept at a temperature of 2300\;K. 
The shape of the tip was observed after every cycle with a Scanning Electron Microscope (SEM) and by applying a smaller cathode field in order to observe its field emission pattern. 
Depending on the local anode field strength (which they denote $F_r$ in \cite{fujita2007mechanism}), they observed different characteristic faceting patterns on the W tip.
The different observed stages of the faceting were as follows \cite{fujita2007mechanism}:
\begin{enumerate}
 \item \label{fujita_A} The original W tip is orientated in the $\langle100\rangle$ direction, relative to the field. It is rounded by a flashing procedure.
 \item \label{fujita_B} With a field of 2.5\;GV/m applied, the the \{110\}, \{211\}, and \{100\} facets became larger in size.
 \item \label{fujita_C} With 3.51\;GV/m, the \{110\}, \{211\}, and \{100\} facets were seen to have grown even more in size and the emitter obtained a polyhedral shape.
 \item \label{fujita_D} With a field of 3.56\;GV/m, the \{211\} facets had suddenly become smaller and the \{110\} facets larger.
 \item \label{fujita_E} With 4.00\;GV/m, there were no \{211\} facet and the \{110\} facets dominated. 
 \item \label{fujita_F} With 4.20\;GV/m, a large square-formed top \{100\} facet is formed with four corners. The field-emission pattern shows four bright spots at the positions of the four corners, where four microprotrusions could be observed with SEM.
 \item \label{fujita_G} With an even larger field of 4.49\;GV/m, an ``overremolding'' state is reached where the surface becomes irregular and the emission pattern is not any more predictable.
\end{enumerate}

\begin{figure*}
 \centering
\subfigure[Initial]{
  \includegraphics[width=0.3\textwidth]{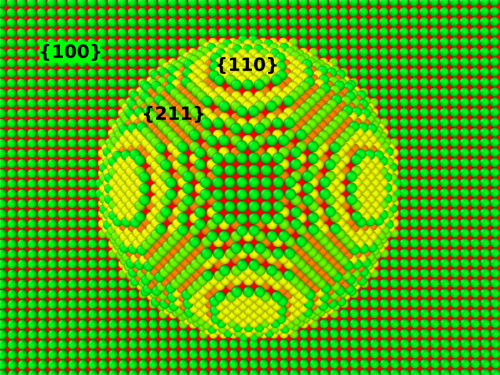}\label{fig:fujita_simulations_a}
}
\subfigure[No field, $t = 20$\;ns]{
  \includegraphics[width=0.3\textwidth]{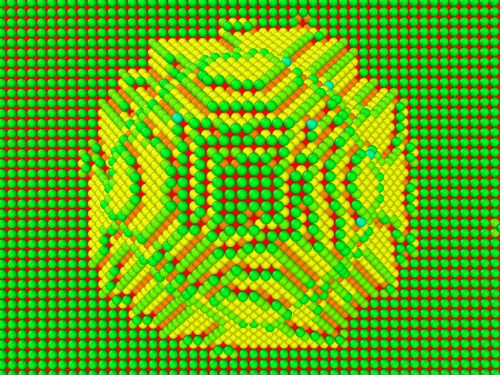}\label{fig:fujita_simulations_b1}
}
\subfigure[No field, $t = 100$\;ns]{
  \includegraphics[width=0.3\textwidth]{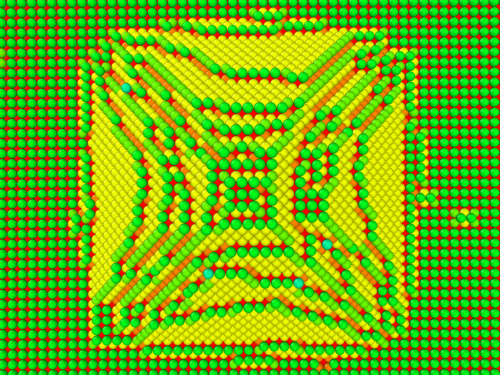}\label{fig:fujita_simulations_b2}
}
\subfigure[$F = 2.5$\;GV/m, $t = 20$\;ns]{ 
  \includegraphics[width=0.3\textwidth]{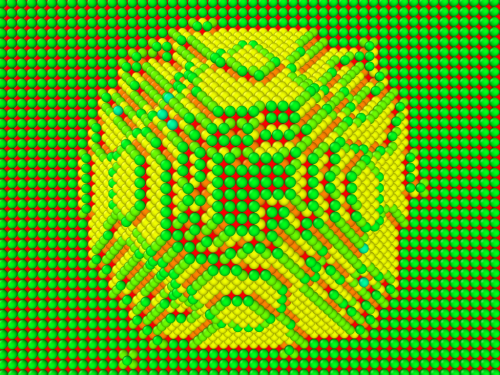}\label{fig:fujita_simulations_c}
}
\subfigure[$F = 20$\;GV/m, $t = 20$\;ns]{ 
  \includegraphics[width=0.3\textwidth]{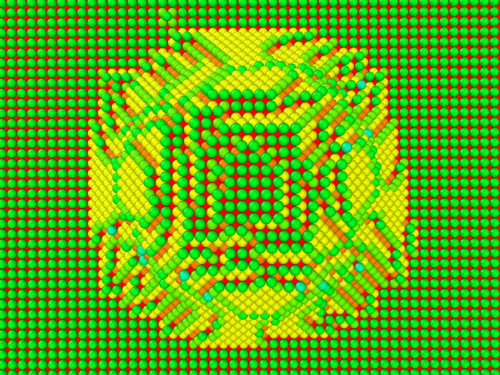}\label{fig:fujita_simulations_d}
}
\subfigure[$F = 30$\;GV/m, $t = 20$\;ns]{ 
  \includegraphics[width=0.3\textwidth]{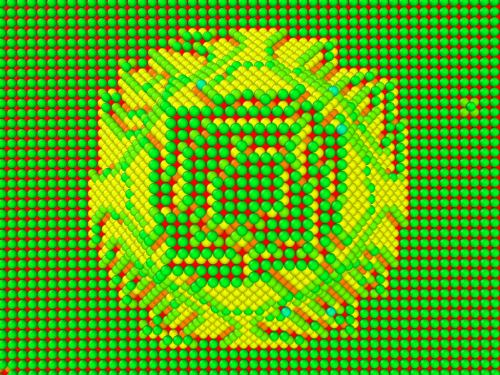}\label{fig:fujita_simulations_e}
}
\subfigure[$F = 50$\;GV/m, $t = 20$\;ns]{ 
  \includegraphics[width=0.3\textwidth]{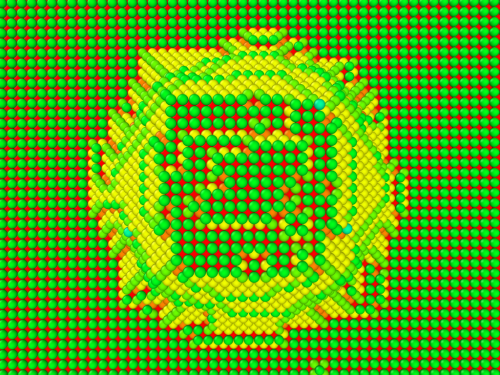}\label{fig:fujita_simulations_f}
}
\subfigure[$F = 60$\;GV/m, $t = 8.6$\;ns ]{ 
  \includegraphics[width=0.3\textwidth]{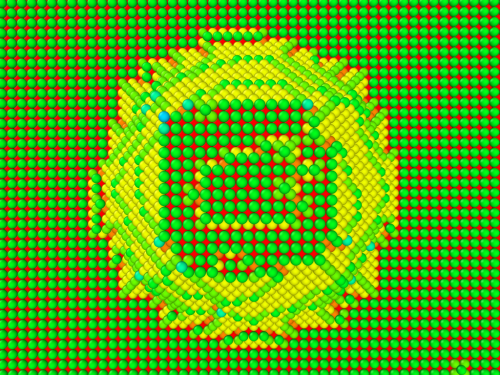}\label{fig:fujita_simulations_g}
}
\subfigure[$F = 60$\;GV/m, $t = 20$\;ns]{ 
  \includegraphics[width=0.3\textwidth]{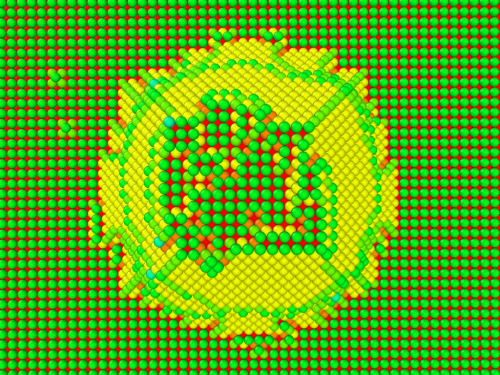}\label{fig:fujita_simulations_h}
}

\caption{Top view of the simulations of the hemispherical tip system (a) at different applied fields. (b)--(c) shows the same simulation of the hemisphere after 20\;ns and 100\;ns, respectively, without any field. In (d)--(g), the system is shown after 20.0\;ns when different applied fields have been applied. (h) and (i) show the same simulation at an applied field of 60\;GV/m after 8.6\;ns and 20\;ns, respectively.
The atoms are coloured according to the coordination numbers to emphasise the different facets. }
\label{fig:fujita_simulations}
\end{figure*}

In order to reproduce with Kimocs these above listed experimentally observed stages of the W tip evolution from \cite{fujita2007mechanism}, we used a system with a smaller 3\;nm high W hemisphere with a 10\;nm radius (figure \ref{fig:fujita_simulations_a}) on a substrate with a perfectly flat W\{100\} surface.
The flat substrate is only needed to create a lower boundary for the system and the field solver; for the comparison with the experiment, we are only interested into which faceting pattern the hemispherical tip evolves at different applied constant fields.

The system was a cubic box with every side of length 64$a_0$, with the W lattice parameter $a_0 = 3.14339\cdot10^{-10}$\;m. 
The substrate was 3$a_0$ (0.94\;nm) thick, leaving 16\;nm of vacuum above the apex of the 3\;nm hemispherical tip on the substrate.
This is more than enough vacuum to make the field homogeneous at the upper boundary of the system.
The thickness of the substrate is not important, since almost all atomic movement will happen on the tip due to the locally enhanced field, whereas the strongly bonded atoms on the perfectly flat surface of the substrate will not move significantly.
Periodic boundaries were applied in the lateral $x$ and $y$ directions. 
The bottom atom layer of the substrate was fixed. 

Different electric fields were applied in the system by solving the Laplace equation in the vacuum part of the system, above the surface, as described in section \ref{sec:field_solver}. 
Thirteen different applied anode fields between 2.5 and 60.0\;GV/m were used, defined as the surface field at the centre of the top \{100\} facet of the initial hemisphere and should be comparable to the applied fields in the experiment of Fujita and Shimoyama.
The corresponding fields at the upper boundary of our vacuum system were in fact between 1.48 and 35.4\; GV/m, respectively.
The simulations were stopped after 20\;ns simulated time had passed, which corresponds to between $7.4\cdot10^5$ and $8.7\cdot10^5$ KMC steps, or about four weeks in CPU time. 
\begin{figure*}
 \centering
\subfigure[Initial]{
  \includegraphics[width=0.3\textwidth]{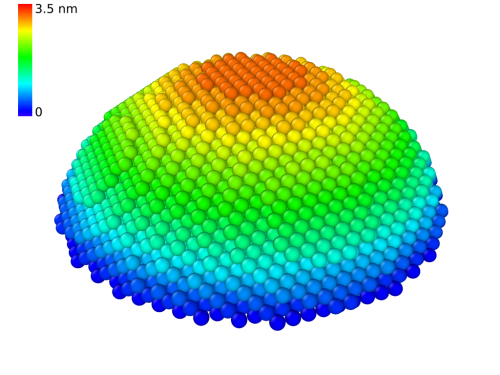}\label{fig:fujita_simulations_perspective_a}
}
\subfigure[No field, $t = 20$\;ns]{
  \includegraphics[width=0.3\textwidth]{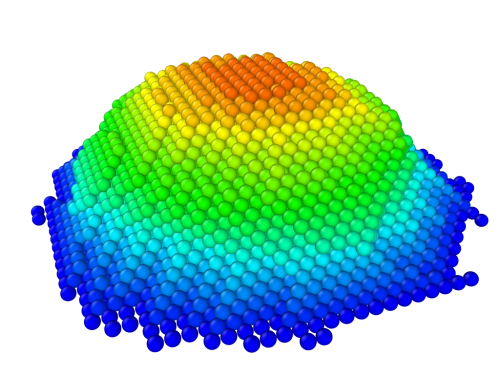}\label{fig:fujita_simulations_perspective_b1}
}
\subfigure[No field, $t = 100$\;ns]{
  \includegraphics[width=0.3\textwidth]{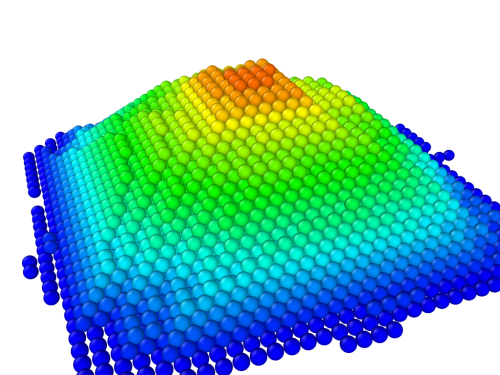}\label{fig:fujita_simulations_perspective_b2}
}
\subfigure[$F = 50$\;GV/m, $t = 20$\;ns]{ 
  \includegraphics[width=0.3\textwidth]{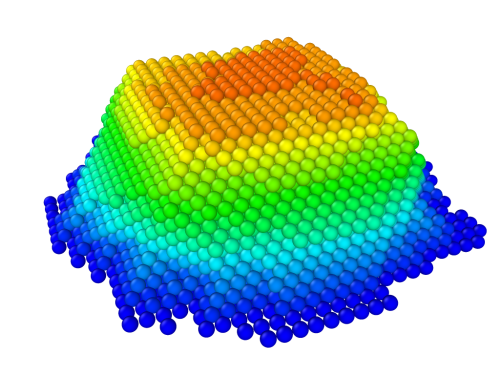}\label{fig:fujita_simulations_perspective_c}
}
\subfigure[$F = 60$ GV/m, $t = 8.6$\;ns]{
  \includegraphics[width=0.3\textwidth]{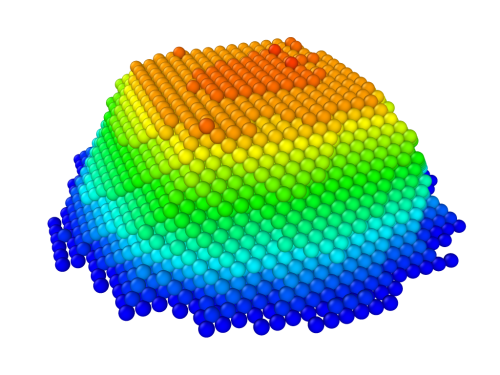}\label{fig:fujita_simulations_perspective_d}
}
\subfigure[$F = 60$ GV/m, $t = 20$\;ns]{
  \includegraphics[width=0.3\textwidth]{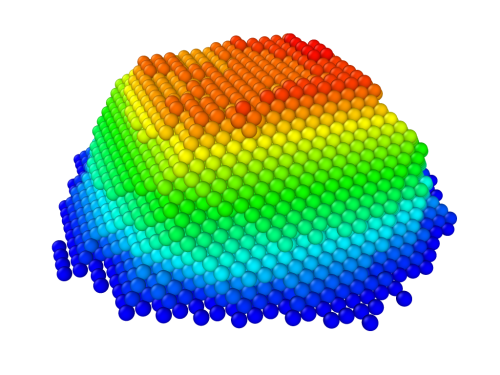}\label{fig:fujita_simulations_perspective_e}
}

\caption{Perspective views of the hemispherical tip: (a) initial, (b)--(c) the same simulations after 20\;ns and 100\;ns, respectively,  without field; (d) with a 50\;GV/m applied field after 20\;ns; (e) and (f) show the same tip with a 60\;GV/m applied field after 8.6\;ns and 20\;ns, respectively.
The atoms are coloured according to the height from the substrate (not shown). The figures are made with Ovito \cite{ovito}.}
\label{fig:fujita_simulations_perspective}
\end{figure*}

The final surfaces after the simulations with the different applied fields are, for selected cases, shown in figure \ref{fig:fujita_simulations} (top views) and figure \ref{fig:fujita_simulations_perspective} (perspective views). 
The initial system is shown in figures \ref{fig:fujita_simulations_a} and \ref{fig:fujita_simulations_perspective_a}. 
As mentioned before, the hemispherical tip has a curvature that, for computational reasons, is $\sim$20 times smaller than the curvature of the tip in Fujita's experiment.
Because the tip is smaller, already the initial tip displays small facets: the top \{100\}, the four \{110\}, and the four \{211\} facets. 
These will be the same facets as observed in the early faceting of the tip in Fujita's experiment (\ref{fujita_B}) because of the identical lattice orientation $\langle 100 \rangle$.

Figures \ref{fig:fujita_simulations_b1}--\ref{fig:fujita_simulations_b2} and \ref{fig:fujita_simulations_perspective_b1}--\ref{fig:fujita_simulations_perspective_b2} show the tip after 20\;ns or 100\;ns without any field for reference.
In the no-field case, it can be seen that the small initial top \{100\} facet has actually decreased in size and the \{211\} facets have grown.

In figure \ref{fig:fujita_simulations_c}, a field of 2.5\;GV/m has been applied and the top \{100\} facet is now more prominent than in the case without field [figure \ref{fig:fujita_simulations_b1}--\ref{fig:fujita_simulations_b2}].
The \{211\} facets are larger than in the initial case [figure \ref{fig:fujita_simulations_a}], but not much different to the no-field case.
Figure \ref{fig:fujita_simulations_c} seems to roughly correspond to Fujita's stages (\ref{fujita_B}) or (\ref{fujita_C}).

With an applied field of 20\;GV/m, shown in figure \ref{fig:fujita_simulations_d}, the \{211\} facets start to decrease and the the \{110\} facets start to grow, corresponding to Fujita's stage (\ref{fujita_D}). 

With an applied field of 30\;GV/m [figure \ref{fig:fujita_simulations_e}], the \{211\} facets have completely disappeared and the \{110\} facets dominates, as also happened in Fujita's stage (\ref{fujita_E}). 

With an applied field of 50\;GV/m, a clear, fairly quadratic, top \{100\} facet is formed, as can be seen in figures \ref{fig:fujita_simulations_f} and \ref{fig:fujita_simulations_perspective_c}.
This facet is formed as atoms along the side \{110\} have a biased diffusion towards the higher fields at the edge of the \{100\} facet. 
However, for an edge atom to overcome the edge onto the \{100\} facet, a barrier between 0.8 and 3.6\;eV needs to be overcome, which is relatively high compare to the 0.6\;eV barrier of the simple jump on the side \{110\} facets.
The end result is that the \{110\} islands migrate towards the \{100\} edge in the same time as they grow in size.
The tip itself will not grow in height (at this stage), but the \{110\} facets will grow sideways until they fill up the corners of the tip, where initially were \{211\} facets, and in the same time growing the corners of the top \{100\} facet.
This is in contrast to the case with no field [figures \ref{fig:fujita_simulations_b1}--\ref{fig:fujita_simulations_b2}], where the \{211\} and \{110\} facets grows and connects with the substrate, where new layers slowly form, making the tip wider. 
The top \{100\} facet, on the other hand, almost completely disappear.

An slightly more pronounced \{100\} facet is also formed with an applied field of 60\;GV/m after 8.6\;ns, as shown in figures \ref{fig:fujita_simulations_g} and \ref{fig:fujita_simulations_perspective_d}.
This is in good agreement with the large \{100\} top facet Fujita and Shimoyama observed in their stage (\ref{fujita_F}), even though we have used higher fields than they reported.
The obtained shape of the tip with the large \{100\} facet is clearly different than the tip shape obtained in the case with no field, figures \ref{fig:fujita_simulations_b1}--\ref{fig:fujita_simulations_b2}, indicating that the obtained evolution is truly an effect of the applied electric field.

Continuing the simulation with the applied field of 60\;GV/m shown in figures \ref{fig:fujita_simulations_g} and \ref{fig:fujita_simulations_perspective_d} at 8.6\;ns, until 20\;ns has passed, shown in figures \ref{fig:fujita_simulations_h} and \ref{fig:fujita_simulations_perspective_e}, no essential change of the tip shape with the prominent top \{100\} facet can be seen, indicating that this shape is indeed very close to the eqilibrium shape.

We did not observe the ``overremolding'' stage (\ref{fujita_G}) that would be expected at fields higher than stage (\ref{fujita_F}), where the large \{100\} facet formed, but overall, the simulated W tips showed qualitatively the same faceting patterns for different electric field strengths, as was observed experimentally in \cite{fujita2007mechanism}. 
This serves as a good validation of the simulation model.


\subsection{Nanotip growth mechanism}\label{sec:nanotip}


In the previous section we saw that the KMC model is able to reproduce the surface evolutions in electric fields as observed in experiments. 
In this section we will turn to the more fundamental question whether fields may cause the formation of nanotips by biasing the surface atom diffusion: what kind of field strengths are needed to promote growth, what is the influence of the temperature, and is there a difference between the influence of anode and cathode fields?

\begin{figure*}[p]
 \centering
 \subfigure[$t = 0$]{
 \includegraphics[width=0.9\columnwidth]{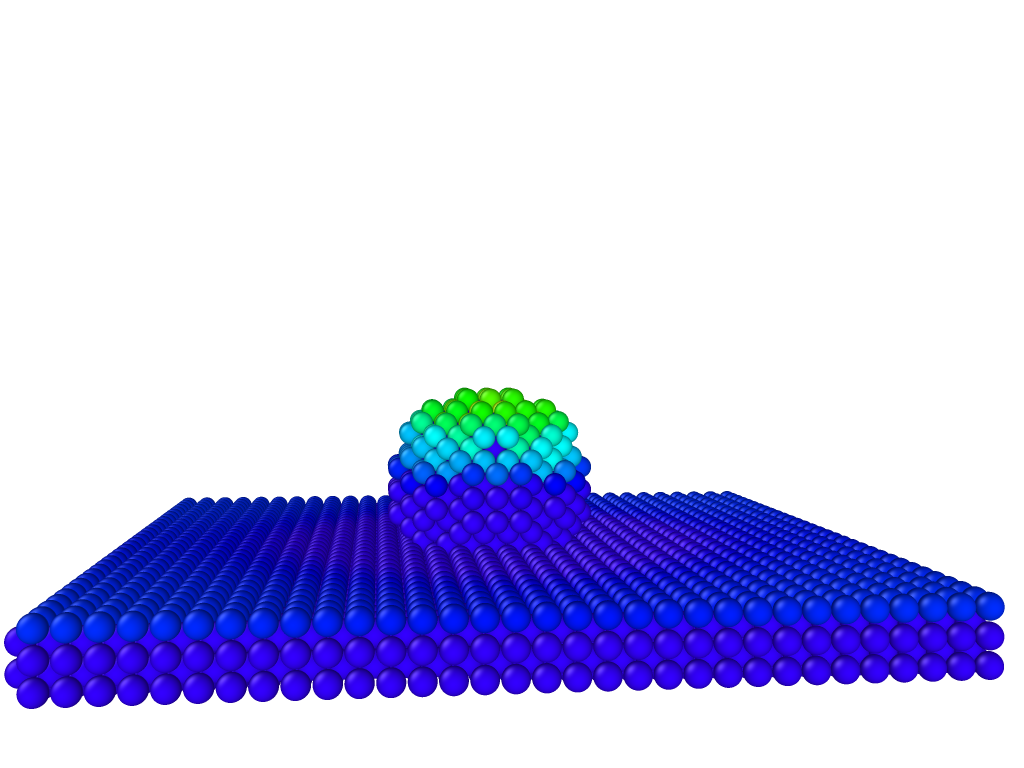}\label{fig:tip_growth_perspective_a}
 }
 \subfigure[$t = 2.0$ ns]{
 \includegraphics[width=0.9\columnwidth]{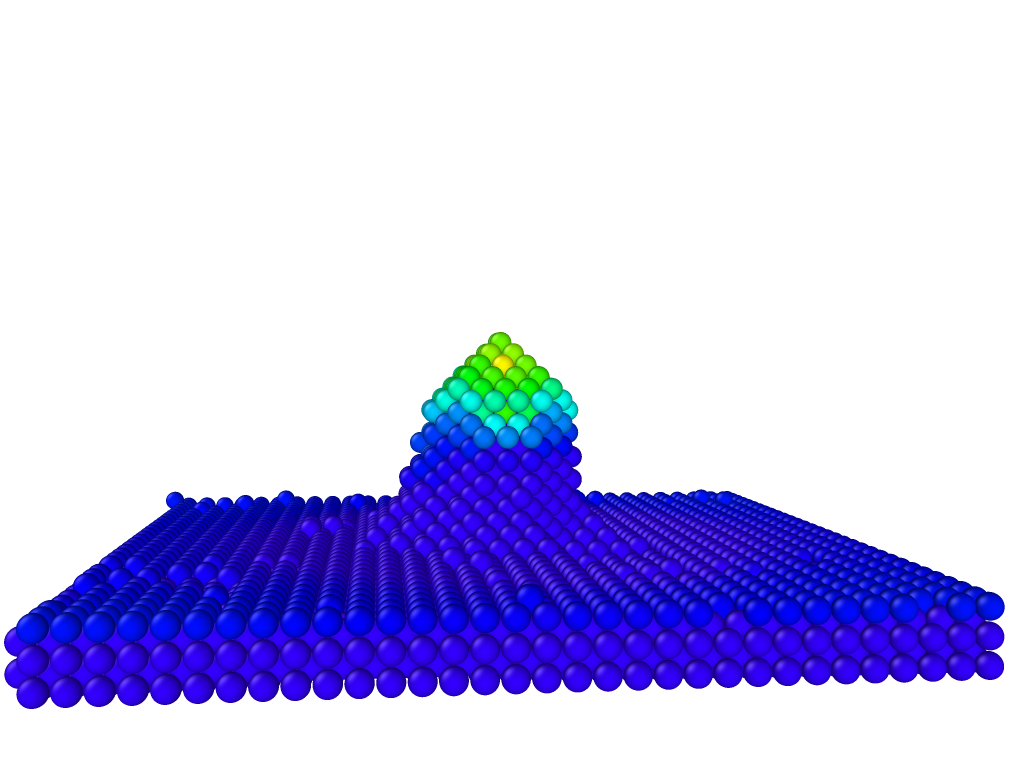}\label{fig:tip_growth_perspective_b}
 }
 \subfigure[$t = 5.0$ ns]{
 \includegraphics[width=0.9\columnwidth]{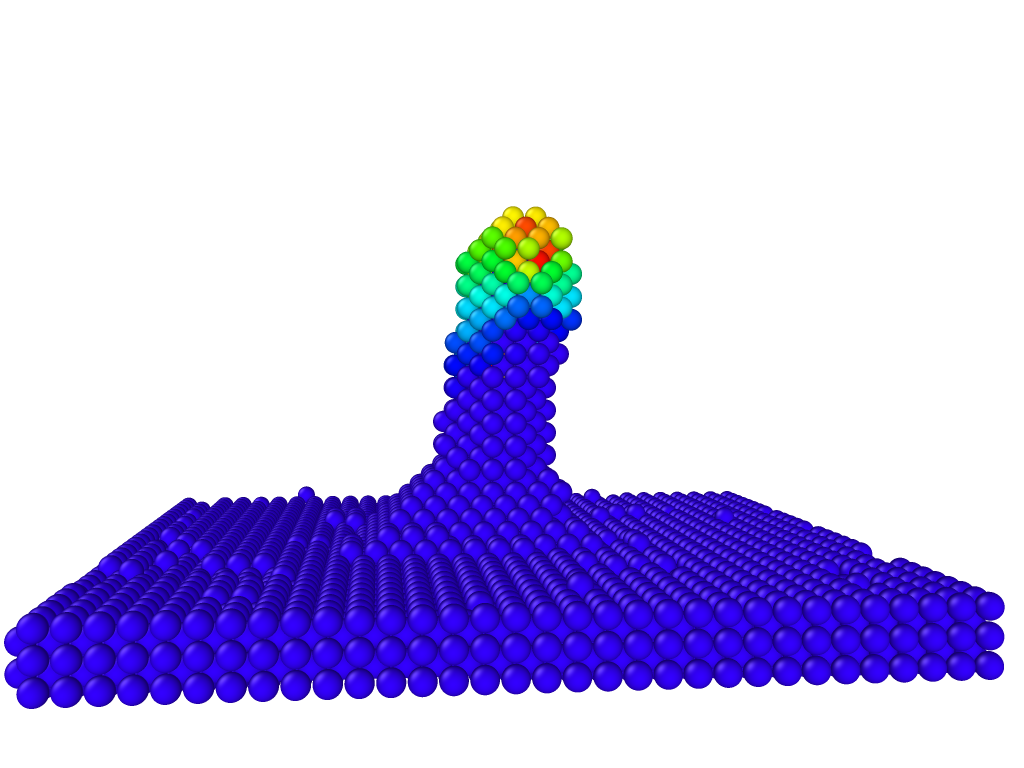}\label{fig:tip_growth_perspective_c}
 }
 \subfigure[$t = 7.6$ ns]{
 \includegraphics[width=0.9\columnwidth]{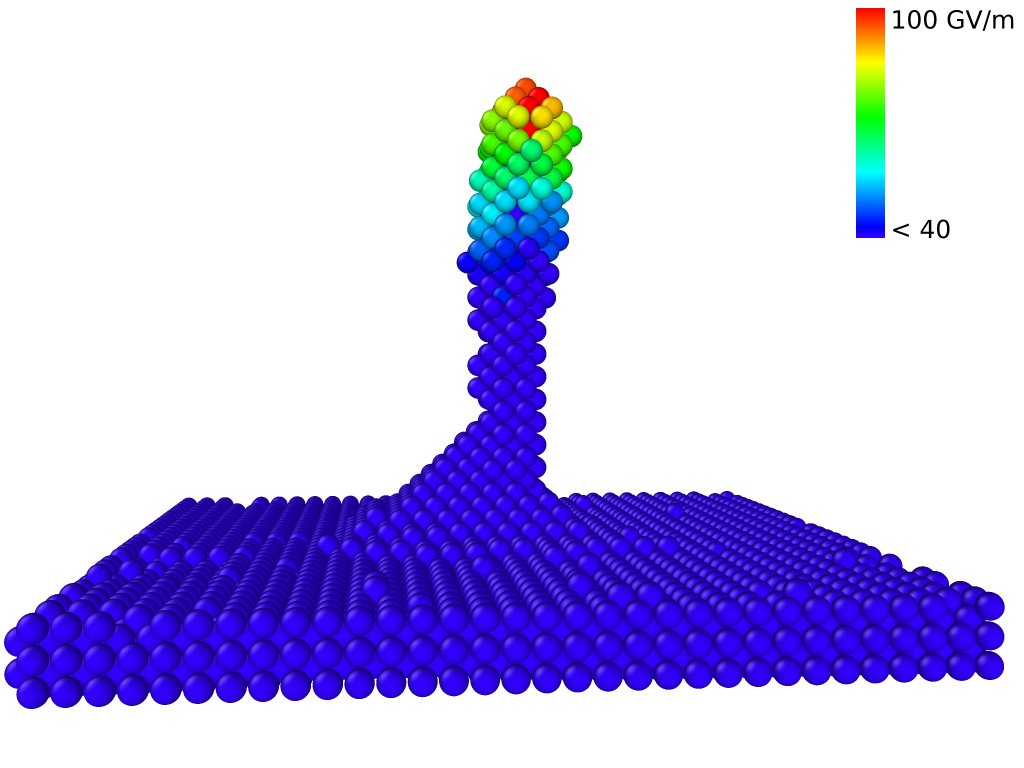}\label{fig:tip_growth_perspective_d}
 }
 \caption{Time frames showing the growth process of a W nanotip at different stages in an applied electric field of 50\;GV/m (initial local field 72\;GV/m) at 3000\;K, starting from a hemispherical asperity (a). The initial asperity (a) is 2.0\;nm high and the final nanotip (d) is 6.3\;nm high.
 The atoms are coloured according to the local surface field strength. 
 For an animation, see the Supplementary Materials.}
 \label{fig:tip_growth_perspective}
\end{figure*}

\begin{figure*}[p]
 \centering
 \subfigure[$t = 0$]{
  \includegraphics[width=0.9\columnwidth]{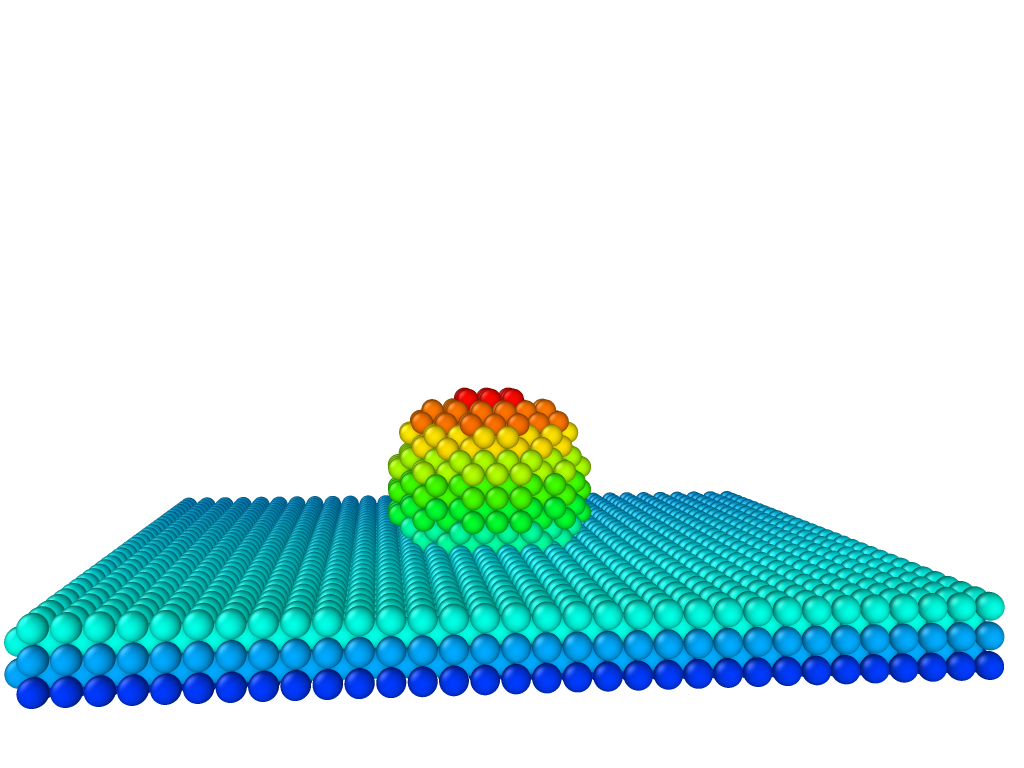}\label{fig:tip_growth_diffusion_a}
 }
  \subfigure[$t = 7.6$ ns]{
 \includegraphics[width=0.9\columnwidth]{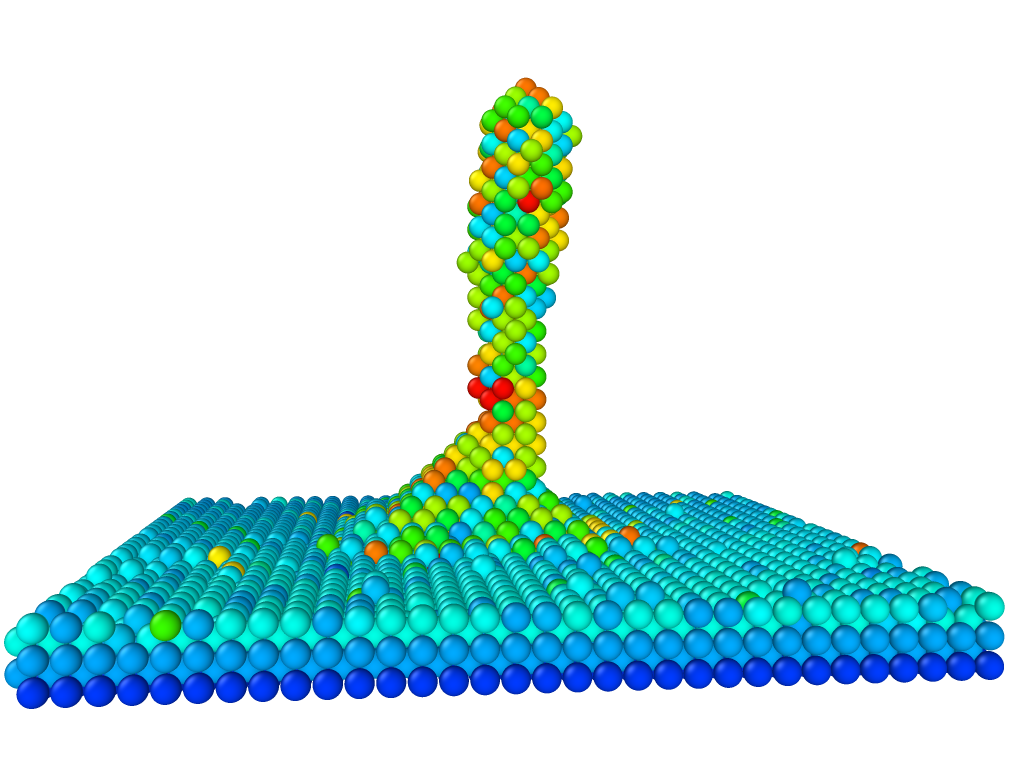}\label{fig:tip_growth_diffusion_b}
 }
 \caption{The same nanotip growth simulation as in figure \ref{fig:tip_growth_perspective}, but with the atoms coloured according to their initial position (a) in order to show how the individual atoms have diffused.}
 \label{fig:tip_growth_diffusion}
\end{figure*}

\begin{figure}
 \centering
 \includegraphics[width=0.99\columnwidth]{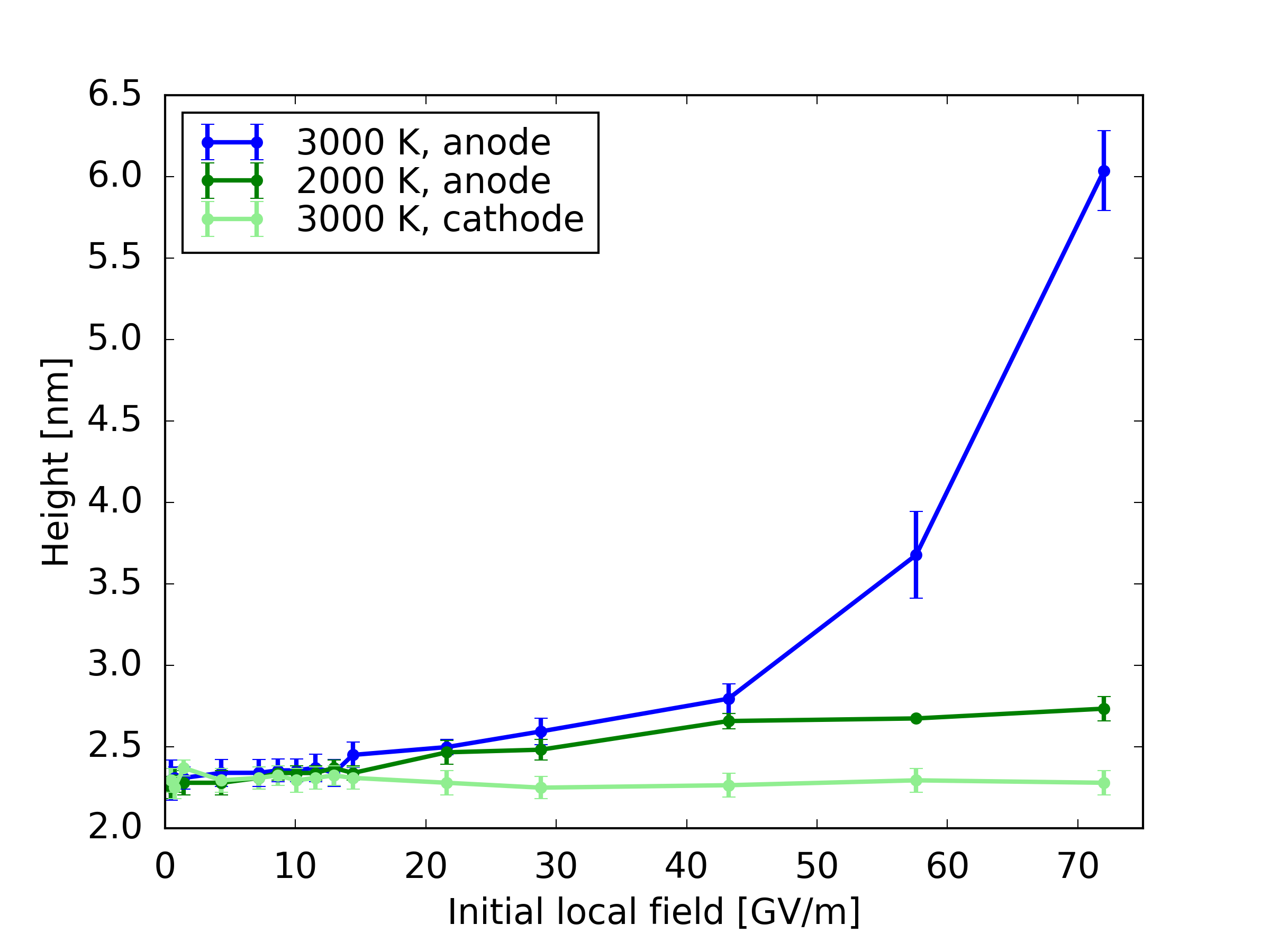}\label{fig:tip_height}
 \caption{Maximum height reached during a $1.5\cdot10^5$ steps simulation (out of ten parallel runs) for different temperatures and initial fields. 
 The initial nanotip height was 2.0\;nm.}
 \label{fig:tip_max_height}
\end{figure}

In these simulations, we used as starting point a smaller hemispherical asperity with a height of 2.0\;nm and a radius of 2.8\;nm, placed on a W\{100\} surface [figure \ref{fig:tip_growth_perspective_a}].
The system was 32$a_0$ wide in the lateral x and y directions and 64$a_0$ in the z direction. 
The substrate was 3$a_0$ thick with the lowest atom layer fixed. 
Periodic boundary conditions were applied in the x and y directions.
Different initial local fields between 0.4 and 72\;GV/m, measured on the top facet of the initial asperity, were used.
These fields corresponds to applied external fields between 0.3 and 50\;GV/m, measured at the upper boundary of the vacuum system.
The locally enhanced fields, measured at a distance $h=1$\;nm above the initial tip, were between 0.4 and 73\;GV/m.
Two different temperatures, 2000\;K and 3000\;K, were used. 
Both are well below the melting temperature of W. 
Corresponding negative applied field values were used for the cathode case, but only one temperature, 3000\;K, was used.
The simulations were stopped after $1.5\cdot10^6$ steps or before that if a maximum height of 6.4\;nm was reached. 
This corresponds to simulated times between 6.9 and 36 ns at 3000\;K and 0.1 to 1.5 µs at 2000\;K.
Generally, longer time scales are reached with lower fields for the same number of steps, since higher fields generally lower the migration barriers and thus shorter the average time steps.
The case with strongest growth, with an initial local anode field $F = 72$\;GV/m at 3000\;K, is shown in figure \ref{fig:tip_growth_perspective}.
The same simulation is also shown in figure \ref{fig:tip_growth_diffusion} where the atoms are coloured according to their original $z$ coordinates, in order to show how individual atoms have diffused.
It can be seen that even atoms from the substrate (blue) have been able to diffuse almost to the apex of the nanotip.
The maximum nanotip height for the different initial local fields and temperatures are shown in figure \ref{fig:tip_max_height}.

The growth is observed to be enhanced by higher temperatures and higher field strengths. 
Some growth is observed at 3000\;K for all applied anode fields, but the growth is more significant above 30\;GV/m. 
The maximum growth, to a height of 6.3\;nm, is found with the highest temperature and field, i.e. at 3000\;K with an initial local field of 72\;GV/m.
This gives a height increase of a factor 3.1.
With cathode fields, at 3000\;K, the initial hemispherical nanotip flattened at all field strengths and no growth was observed.

\subsection{Sensitivity of the model parameters}\label{sec:sensitivity}

To check the sensitivity of the four electronic $\mathcal{M}$ and $\mathcal{A}$ parameters, the best case of the same system as in section \ref{sec:nanotip}, with 3000\;K and an initial local field of 72 GV/m, was rerun with the electronic parameters varied separately by $\pm10$ \% or $\pm 20$ \% to see how the nanotip growth is affected. 

\begin{figure*}
  \centering
  \subfigure[]{
  \includegraphics[width=0.99\columnwidth]{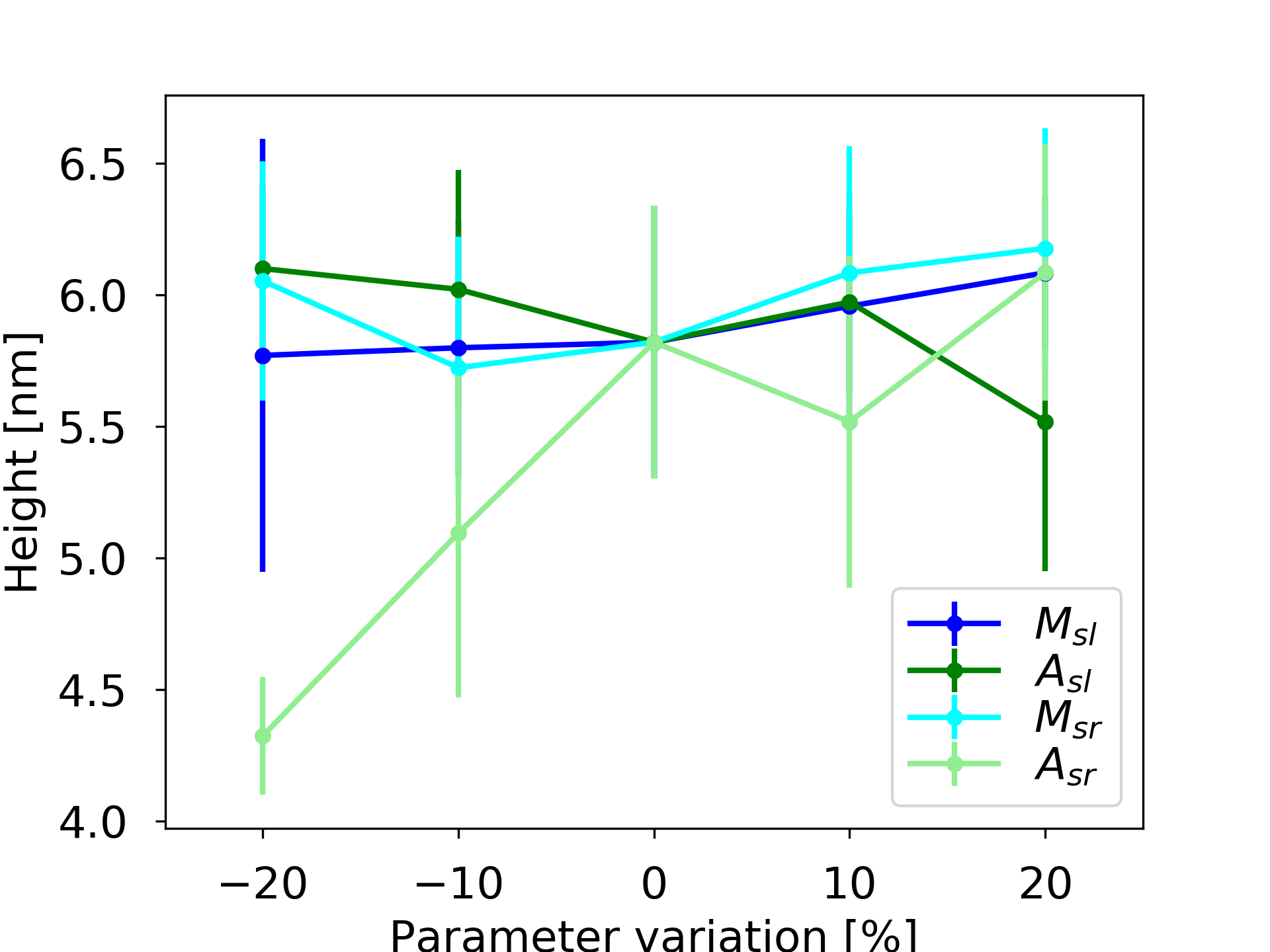}\label{fig:tip_electric}
  }
  \subfigure[]{
  \includegraphics[width=0.99\columnwidth]{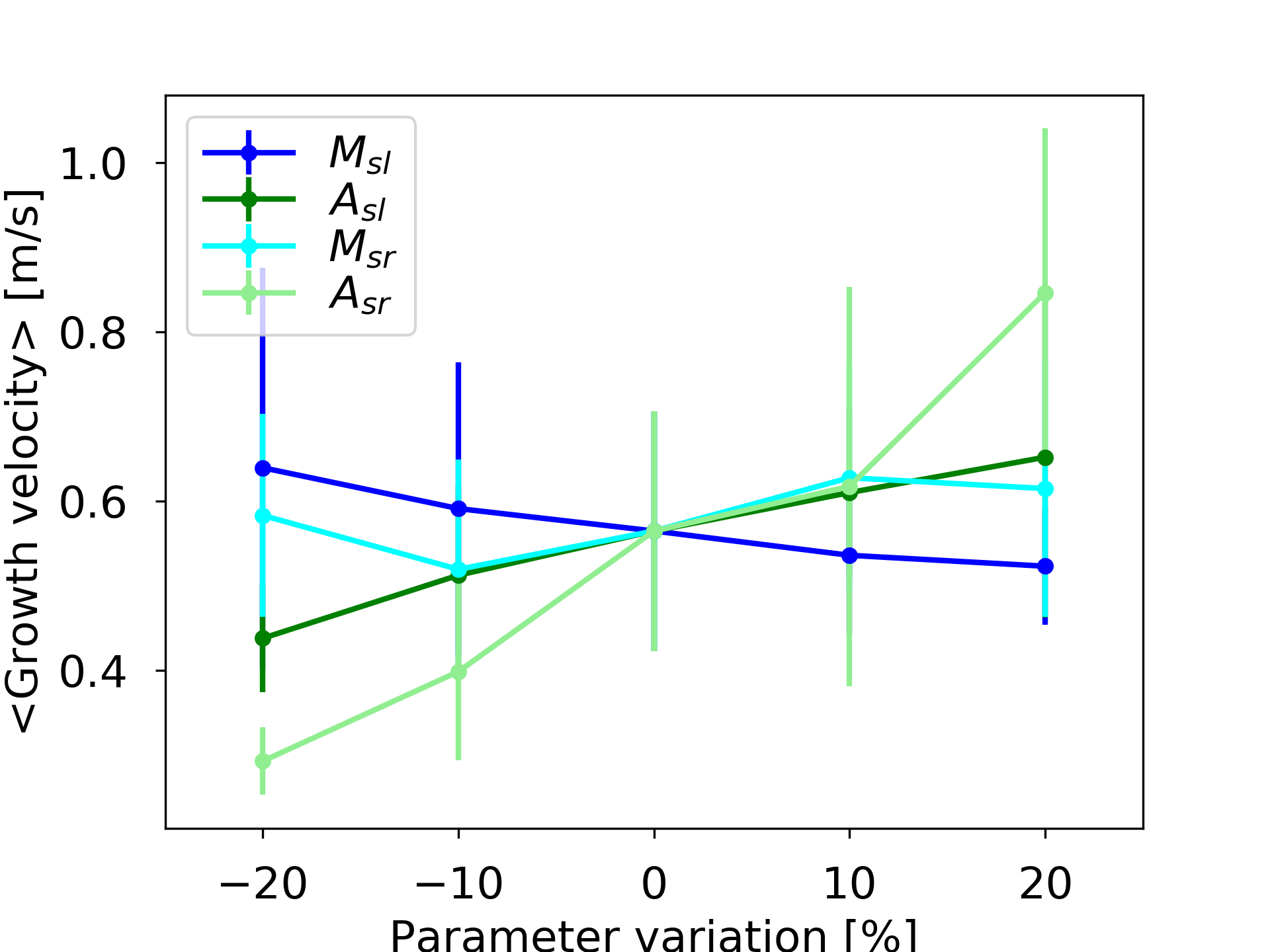}\label{fig:tip_electric_velocity}
  }
  \caption{Sensitivity of the electronic $\mathcal{M}$ and $\mathcal{A}$ parameters. The best case from section \ref{sec:nanotip} is simulated with each of the four $\mathcal{M}$ and $\mathcal{A}$ parameters separately varied by $\pm10\%$ or $\pm20\%$. The reference case is included as 0 \%. Figure (a) shows the nanotip average maximum heights over ten runs, (b) shows the average growth velocity to reach the maximum height. Error bars correspond to the standard deviation.}
  \label{fig:tip_sensitivity}
  
\end{figure*}

In figure \ref{fig:tip_electric} is shown the maximum nanotip height (out of 10 statistical runs) for different variations of the electronic parameters.
The reference case (0\;\%) is done with 50 statistical runs.
It can be seen that even if one of the parameters are changed by 20 \%, the nanotip will still grow significantly at this field and temperature.
The most sensitive parameter is $A_{sr}$ which, when reduced by 20 \%, only grows by a factor $\sim$2.2 to 4.3\;nm, which is still a significant growth.
If any of the other parameters are varied, a height between 5.5 and 6.2\;nm is reached, which is not significantly different from the height of the reference, $(5.8\pm0.5)$\;nm.
If any of the parameters are individually varied by only $\pm10\%$, the reached average maximum heights are all within the even smaller span of 5.0 and 6.1\;nm. 

To better get the effect of the individual $\mathcal{M}$ and $\mathcal{A}$ parameters, the average growth velocity of the tip, defined as the difference in nanotip height divided by the time it took to reach the maximum height, is shown in figure \ref{fig:tip_electric_velocity}.
It can be seen that increasing $\mathcal{M}_{sl}$ decreases the growth velocity, which is the opposite trend as given by varying $\mathcal{A}_{sl}$ or $\mathcal{A}_{sr}$. 
Varying $\mathcal{M}_{sl}$ gives no clear trend in either direction.

We also checked the sensitivity of the $h$ parameter, which gives the distance above the surface where the surface fields $F_l$ and $F_v$ are calculated by the field solver for every atom and surface vacancy.
The same system as in section \ref{sec:nanotip} was again simulated at 3000\;K at external fields between 8.0 and 50\;GV/m, but with $h$ varied between 0.5 and 1.2\;nm.
Changing $h$ will change the values of the fields $F_l$, $F_v$ and the gradients $\gamma$.
The local fields will subsequently also be different for the same external fields, as measured at the upper boundary of the vacuum.

\begin{figure}
  \centering
  \includegraphics[width=0.99\columnwidth]{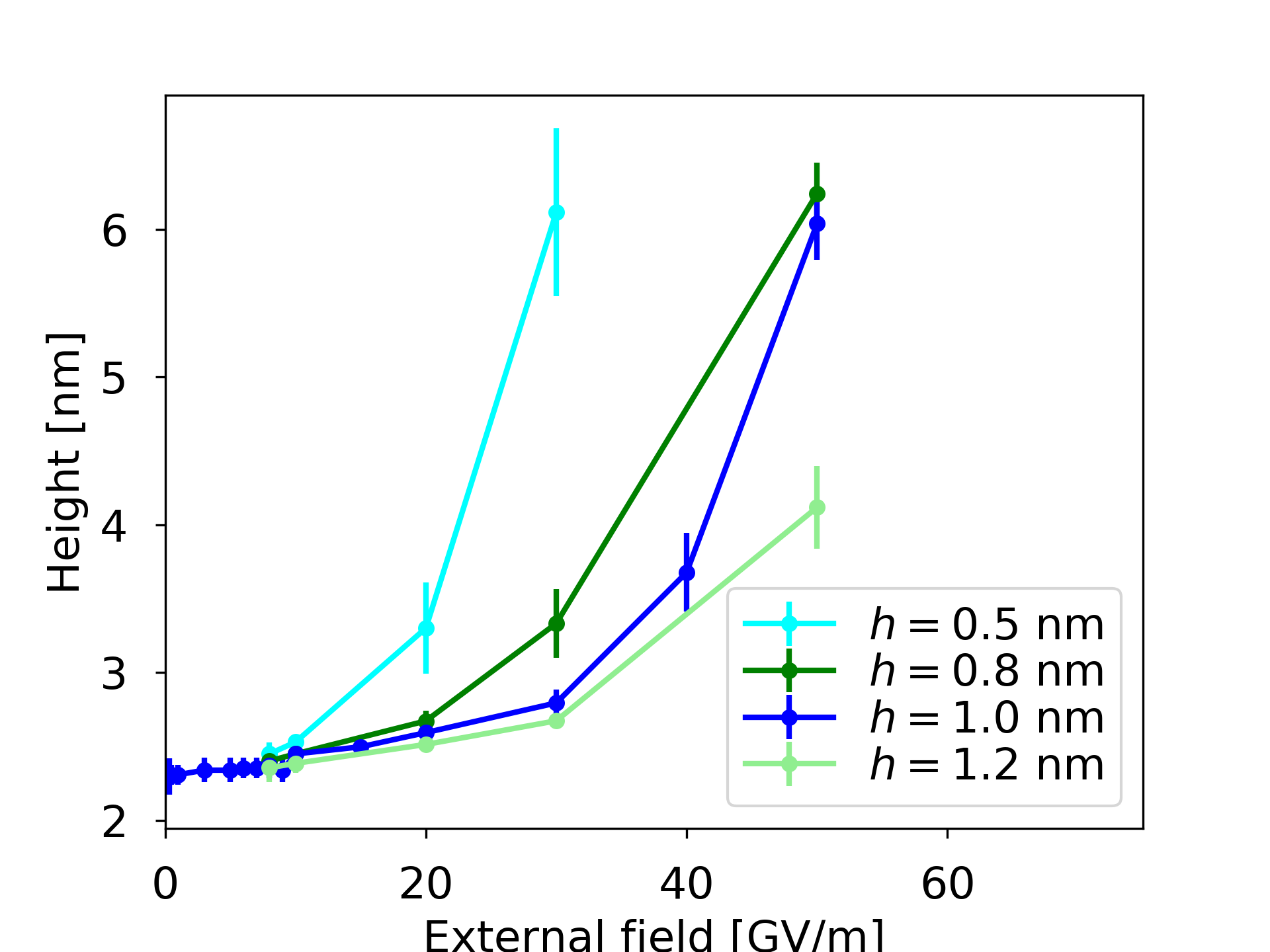}
  \caption{Sensitivity of $h$, the distance above the surface where the surface fields $F_l$ and $F_v$ are calculated by the field solver. The maximum nanotip height is shown for different external fields (measured at the upper vacuum boundary of the simulation box). As standard $h=1.0$\;nm is used. Every data point is the average of ten runs.}
  \label{fig:h_sensitivity}
 \end{figure}

As is shown in figure \ref{fig:h_sensitivity}, with smaller $h$ values, the growth is enhanced and more significant at smaller external fields.
With $h=0.5$\;nm, significant growth appears already at external fields between 1 and 20 GV/m, which corresponds with this $h$ to the initial local fields between 18 and 37 GV/m. 
With $h=0.5$\;nm and an external field of 50\;GV/m, too many atoms started to evaporate (see section \ref{sec:methods}) for the results to be reliable.

\section{Discussion}\label{sec:discussion}

The KMC code Kimocs and its $(a,b,c,d)$ parameterization model for the atom transitions were originally developed for studying only thermally activated diffusion processes on surfaces and has been successfully used in several such studies \cite{jansson2016long,zhao2016formation,vigonski2018au,baibuz2018migration}, where the simulation results have shown good agreement with both MD simulations and experimental results.
The W parameterization in this work \cite{jansson2020tungsten} includes not only first-nearest neighbour atom jumps, but also second-nearest neighbour jumps.
The second-nearest jumps are particularly important on the \{100\} surface, as no first-nearest neighbour jumps are possible there.
Additionally, the parameterization includes a third-nearest neighbour exchange process, as its barrier 1.95\;eV is comparable to the other jumps and therefore may be of significance.
The W parameterization has already been discussed in detail and validated in a previous paper \cite{jansson2020tungsten}.

\subsection{The field model}\label{sec:discussion_model}

In order to extend the KMC model to include electric field, we have used the recently developed theory for how an applied field affects the atom migration barriers \cite{kyritsakis2019atomistic}, summarized in section \ref{sec:theory}.
The electric field is solved in the vacuum space above the atomic surface using the established field solver from {\sc Helmod} \cite{djurabekova2011atomistic}, which has been previously used in several Molecular Dynamics studies where a field was present, e.g. \cite{parviainen2011electronic,parviainen2011molecular}. 
The effect of the electric field on the migration energy barriers is given by equation \ref{eq:barrier_gradient}, described in section \ref{sec:theory}. 
The full theoretical derivation of equation \ref{eq:barrier_gradient} is given in \cite{kyritsakis2019atomistic}.

Some approximations were unavoidable in the model.
In order to obtain the precise field-dependent migration energy barrier $E_m(F)$ for every possible surface atom jump, the parameters $\mathcal{M}_\textrm{sl}$, $\mathcal{A}_\textrm{sl}$, $\mathcal{M}_\textrm{sr}$, and $\mathcal{A}_\textrm{sr}$ (henceforth simply the $\mathcal{M}$ and $\mathcal{A}$ parameters) would ideally need to be calculated with DFT for every such process (there is no field present in bulk processes). 
Since 5--10 separate DFT calculations are needed to obtain the $\mathcal{M}$ and $\mathcal{A}$ parameters for a single atom transition process and there are 100--1000 possible surface atom transition processes, a number of 500--$10^4$ DFT calculations would be needed to obtain a complete set of the $\mathcal{M}$ and $\mathcal{A}$ parameters. 
This is beyond the scope of this work. 
Instead we make the assumption that the $\mathcal{M}$ and $\mathcal{A}$ values obtained for the single process of a W atom jump on a W\{110\} surface in \cite{kyritsakis2019atomistic} (table \ref{table:field_parameters}) are approximately correct for any W atom transition on a W surface.
This assumption is supported by the good agreement between our KMC simulations and the experimental observation of the tungsten surface evolution in fields, as presented in section \ref{sec:fujita}, but also by the low sensitivity that the nanotip growth simulations showed with regards to the $\mathcal{M}$ and $\mathcal{A}$ parameters in section \ref{sec:sensitivity}.
These results will be further discussed in the following sections \ref{sec:discussion_fuijta} and \ref{sec:discussion_nanotip}.

Another uncertainty in the model is how to define and calculate the fields associated with the lattice points $F_l$, the difference in field between the saddle point ($F_s$), and the lattice point, $\Delta F = F_s - F_l$, (as used in equation \ref{eq:barrier_gradient}).
As was already described in section \ref{sec:implementation}, $F_l$, $F_s$, and $\Delta F$ depend on the parameter $h$, which is the distance above the surface, along the field lines, that the field for a particular surface atom or surface vacancy is calculated by the field-solver.
The distance $h$ needs to be large enough for the field not to be affected by the local field enhancement of the adatom itself, as this enhancement is already included in the $\mathcal{A}$ and $\mathcal{M}$ parameters \cite{kyritsakis2019atomistic}.
If, on the other hand, $h$ would be too large, $F_l$ and $F_v$ would both approach the nominal external field of the system, i.e. the limit at the upper boundary of the vacuum, and the gradient $\gamma$ and $\Delta F$ would approach zero.
In section \ref{sec:implementation} we showed for a flat surface with a field gradient that with our chosen value $h=1$\;nm, the error for the fields $F_l$ and $F_v$ are 3.4 \% and for the gradient 13 \%.
In section \ref{sec:sensitivity}, we showed that nanotip growth happens even if $h$ is varied by $\pm20$ \%, with the growth enhanced with smaller $h$.

The described choices for how $F_l$ and $\Delta F$ are calculated have the advantage of being easily applicable to any arbitrarily rough surface.
The easy applicability is important since $F_l$ and $\Delta F$ have to be recalculated for every atom and every possible atom jump in the system, respectively, after every time the field-solver has been called, i.e. at every KMC step.

\subsection{Model validation: faceting in fields}\label{sec:discussion_fuijta}

In section \ref{sec:fujita}, we showed that our KMC model is able to qualitatively reproduce the different main characteristic faceting patterning stages of the surface evolution of a W tip at different electric field strengths, as observed in experiments by S\;Fuijita and H\;Shimoyama \cite{fujita2007mechanism}. 
In the experiment, a single tip was observed under fields that were increased in a stepwise manner, with exposure durations of $\sim$30\;s at each step, between the observations.
The authors assume that the exposure time is enough for the tip to change into a shape that minimizes the surface energy, which will be affected by the applied constant electric field.
Based on this assumption, the strength of the applied field will be the most important factor for determining the final energetically favourable shape and the initial tip shape plays a minor role.

In our simulations, we do not stepwise increase the field strength, but start directly with the intended field strength from the perfect hemispherical tip and run it long enough for the tip shape to reach an equilibrium shape.
If comparing the tip in figures \ref{fig:fujita_simulations_perspective_d} and \ref{fig:fujita_simulations_perspective_e}, it can be seen that the tip hardly changes even if the simulation time is doubled.
It is worth adding that if no field applied (figures \ref{fig:fujita_simulations_perspective_b1} and \ref{fig:fujita_simulations_perspective_b2}), the tip will not reach an equilibrium shape but would likely continue to flatten down until the atoms evenly covers the flat substrate, as has been seen in other metal tip simulations \cite{jansson2016long,vigonski2018au,lahtinen2018artificial}.

As we repeated the simulations for all considered field strengths, we could see that the final tip shapes changed with the increased applied electric fields and we could recognize the different distinct faceting stages with increasing fields, as was observed in \cite{fujita2007mechanism}, although our model seems to require much higher fields.
For computational reasons, we were forced to use a tip with much smaller curvature (10\;nm) than the experimental tip ($\sim$206\;nm), which will also change the surface field gradients and the local field enhancements on the surface.
Calculating the field in the simulations is very time-consuming and already with this 10\;nm tip, each simulation took around four weeks to finish.

Fujita and Shimoyama observe the large top \{100\} facet to form at an estimated average field of 4.20\;GV/m, whereas, in the simulations, a similar facet is observed [figures \ref{fig:fujita_simulations_f} and \ref{fig:fujita_simulations_g}] to form at a higher field of 50.0\;GV/m; a difference of a factor $\sim$12. 
This difference is overestimated by the fact that in Ref. \cite{fujita2007mechanism} the local electric field was estimated using the same initial voltage conversion factor (denoted $kr$ in \cite{fujita2007mechanism}) for the entire remolding process, which is not accurate since from Fujita's Fig. 4a, the extraction voltage for the same field emitted current drops by a factor of about 2.5 during the remolding process due to the tip sharpening.
This implies that the conversion length $kr$, i.e. the ratio between voltage and local field is also reduced (the field enhancement increased) by a similar factor.
Therefore a more accurate estimation would be that the local field at the tip is about 11 \;GV/m in that phase of the experiment, which reduces the discrepancy to a factor of $\sim$4.5.

Finally, in figure \ref{fig:fujita_simulations_perspective}, it is clearly shown that if a field of 50--60\;GV/m, the hemispherical tip will develop a characteristic large top \{100\} facet, which was also observed by Fujita and Shimoyama in their stage (\ref{fujita_F})\cite{fujita2007mechanism}.
If, on the other hand, no field is applied, the hemispherical tip will instead evolve into a pyramid shape and never grow any \{100\} facets.
This large top \{100\} facet is of particular importance, as it is a clear field effect and reproduced by the model in good agreement with experiments.
This, together with the observations of the other faceting stages, is a strong validation of the simulation model.

\subsection{Nanotip growth mechanism}\label{sec:discussion_nanotip}

In section \ref{sec:nanotip}, we demonstrated that the migration bias given by surface field gradients may create a nanotip growth mechanism, given that the applied fields and the temperatures are high enough.
Our simulations show that a 2\;nm high hemispherical W tip with a 2.8\;nm curvature at 3000\;K starts to show significant growth at an initial anode field of 9\;GV/m at the tip apex, whereas at 2000\;K, the growth speed is significantly reduced and only some minimal growth can be seen within the simulated time frame of $\sim$1\;µs.
For cathode fields, we do not not see any clear growth with the current model.
The dependence of the sign of the field (anode or cathode) comes from the $\mathcal{M}$ and $\mathcal{A}$ parameters in equation \ref{eq:barrier_gradient}, which are dependent on the electrical properties on the material; in this case tungsten (see also the discussion in section \ref{sec:discussion_model}). 

The simulations of the small tip in section \ref{sec:nanotip} and the larger curved surface in section \ref{fig:field_system} show that the growth processes will be initiated where the field gradients are strongest, such as at the tip apexes or at edges. 
Elsewhere, such as on the flat substrate or at the base of the tips, where the fields are also weaker, only minimal changes of the surface can be seen.
Thus, in order to initiate growth of a nanotip, a nucleation point in the form of some kind of asperity with a high enough gradient is needed.

In this work, we are studying the diffusion growth mechanism, but it is worth noting that we do not take into account thermal and ionic evaporation processes that would give the opposite effect.
These kind of processes may play a significant role at fields 30--50\;GV/m and above.
Atoms at the tip apex may be evaporated if the local fields are strong enough and thereby reduce the tip growth.
However, a precise description of the competing thermal and field-assisted atom evaporation processes is beyond the scope of this work.

We also note that this work only considers nanotips on the bcc\{100\} surface, as the field solver currently only can handle this particular surface.
In \cite{jansson2016long} different stabilities of nanotips of different lattice orientations were found.
Tips on other surfaces than \{100\} may therefore behave and grow differently than observed in this work.

In general, our results make it highly plausible that nanotips may form on metallic surfaces if a high electric field gradient and some kind of nucleation point are present.
Such nanotip growth may be one explanation for the formation of vacuum arcs (or breakdowns) in high electric fields, such as in accelerating structures \cite{navitski2013field,nagaoka2001field,engelberg2018stochastic,clic2018compact}.
The nucleation points, small asperities on the surface, may possibly be caused by dislocation movement close to the surface and dislocation activity in surfaces under high fields has been correlated to breakdowns in a recent statistical model \cite{engelberg2018stochastic}.
Very large nanotips ($\sim$90\;nm high) may, on the other hand, emit enough electrons and neutrals to initiate a vacuum arc plasma \cite{kyritsakis2018thermal}.
Our work provides a plausible mechanism for how these large nanotip may form, where small protrusions, caused by dislocations, are acting as nucleation points that triggers the biased surface atom diffusion in the high fields to build up nanotips large enough to eventually start field emissions and trigger the vacuum arcs.

\section{Conclusions}\label{sec:conclusions}

In this work, we have identified a mechanism where nanotips may start to form on metal surfaces due to biased atom diffusion in high electric fields.

For this work, we have developed a general-purpose Kinetic Monte Carlo model for simulations of diffusion processes on arbitrarily rough tungsten (W) surfaces in high electric fields.
The model was validated by reproducing the experimentally observed stages of the characteristic field-induced faceting patterns of a hemispherical tungsten surface. 
In particular, the formation of a large top \{100\} facet, which only happened under the influence of an applied electric field, was reproduced in good agreement with experiments.

We investigated the nanotip growth mechanism by simulating how small W surface asperities would grow due to the biased atom diffusion into nanotips at different electric field strengths and different temperatures.
We found that the diffusion-driven growth is more significant with higher local fields and higher temperatures.
The model clearly shows the plausibility of the hypothesis that nanotips may form in high electric fields and the results thus also support the hypothesis that vacuum arcs in high-field environments may be caused by such spontaneous formations of nanotips.

\ack
V\;Jansson was supported by Academy of Finland (Grant No.\;285382) and Waldemar von Frenckells Stiftelse. 
A\;Kyritsakis were supported by a CERN K-contract (No.\;47207461).
E\;Baibuz was supported by the CERN K-contract (No.\;47207461) and the doctoral school DONASCI of University of Helsinki.
The work of V\;Zadin and A\;Aabloo was supported by Estonian Research Council Grants PUT1372 and IUT20-24.
F\;Djurabekova acknowledges gratefully the financial support of Academy of Finland (Grant No.\;269696).
Computing resources were provided by the Finnish IT Center for Science (CSC) (persistent identifier urn:nbn:fi:research-infras-2016072533).

\section*{ORCID iDs}
{\small
V\;Jansson     \url{https://orcid.org/0000-0001-6560-9982}\\
E\;Baibuz      \url{https://orcid.org/0000-0002-9099-1455}\\
A\;Kyritsakis  \url{https://orcid.org/0000-0002-4334-5450}\\
S\;Vigonski    \url{https://orcid.org/0000-0002-2849-2882}\\
V\;Zadin       \url{https://orcid.org/0000-0003-0590-2583}\\
S\;Parviainen  \url{https://orcid.org/0000-0003-4571-4640}\\
A\;Aabloo      \url{https://orcid.org/0000-0002-0183-1282}\\
F\;Djurabekova \url{https://orcid.org/0000-0002-5828-200X}
}

\bibliographystyle{my-iopart-num.bst}
\bibliography{pub/vjansson.bib,pub/vjansson_publications.bib}

\providecommand{\newblock}{}
\providecommand{\url}[1]{{\tt #1}}
\providecommand{\urlprefix}{}
\providecommand{\href}[2]{#2}
\begin{thebibliography}{10}

\bibitem{stroscio1991atomic}
Stroscio J~A and Eigler D 1991 {\em Science\/} {\bf 254} 1319--26

\bibitem{tsong2005mechanisms}
Tsong T~T 2005 {\em Physica A: Statistical Mechanics and its Applications\/}
  {\bf 357} 250--81

\bibitem{parviainen2011electronic}
Parviainen S, Djurabekova F, Timko H and Nordlund K 2011 {\em Computational
  Materials Science\/} {\bf 50} 2075--9

\bibitem{kyritsakis2018thermal}
Kyritsakis A, Veske M, Eimre K, Zadin V and Djurabekova F 2018 {\em Journal of
  Physics D: Applied Physics\/} {\bf 51} 225203

\bibitem{de2018electric}
de~Knoop L, Kuisma M~J, L{\"o}fgren J, Lodewijks K, Thuvander M, Erhart P,
  Dmitriev A and Olsson E 2018 {\em Physical Review Materials\/} {\bf 2} 085006

\bibitem{de2019electric}
de~Knoop L, Kuisma M~J, L{\"o}fgren J, Lodewijks K, Thuvander M, Erhart P,
  Dmitriev A and Olsson E 2019 {\em Microscopy and Microanalysis\/} {\bf 25}
  1830--1

\bibitem{kyritsakis2019atomistic}
Kyritsakis A, Baibuz E, Jansson V and Djurabekova F 2019
  \href{https://doi.org/10.1103/PhysRevB.99.205418}{{\em Phys. Rev. B\/} {\bf
  99}(20) 205418} (\textit{Preprint}
  \href{https://arxiv.org/abs/1808.07782}{{\tt 1808.07782}})
  \urlprefix\url{https://doi.org/10.1103/PhysRevB.99.205418}

\bibitem{whitman1991manipulation}
Whitman L, Stroscio J~A, Dragoset R~A and Celotta R 1991 {\em Science\/} {\bf
  251} 1206--10

\bibitem{mendez1996diffusion}
Mendez J, G{\'o}mez-Herrero J, Pascual J, Saenz J, Soler J and Baro A 1996 {\em
  Journal of Vacuum Science \& Technology B: Microelectronics and Nanometer
  Structures Processing, Measurement, and Phenomena\/} {\bf 14} 1145--8

\bibitem{mayer1999electric}
Mayer T, Houston J, Franklin G, Erchak A and Michalske T 1999 {\em Journal of
  applied physics\/} {\bf 85} 8170--7

\bibitem{dulot2000stm}
Dulot F, Eugene J, Kierren B and Malterre D 2000 {\em Applied surface
  science\/} {\bf 162} 86--93

\bibitem{bettler1960activation}
Bettler P~C and Charbonnier F~M 1960
  \href{https://doi.org/10.1103/PhysRev.119.85}{{\em Phys. Rev.\/} {\bf 119}(1)
  85--93} \urlprefix\url{https://link.aps.org/doi/10.1103/PhysRev.119.85}

\bibitem{dyke1960electrical}
Dyke W~P, Charbonnier F~M, Strayer R~W, Floyd R~L, Barbour J~P and Trolan J~K
  1960 \href{https://doi.org/10.1063/1.1735700}{{\em Journal of Applied
  Physics\/} {\bf 31} 790--805}

\bibitem{zhou2005growth}
Zhou J, Gong L, Deng S~Z, Chen J, She J~C, Xu N~S, Yang R and Wang Z~L 2005
  {\em Applied Physics Letters\/} {\bf 87} 223108

\bibitem{bormann2015ultrafast}
Bormann R, Strauch S, Sch{\"a}fer S and Ropers C 2015 {\em Journal of Applied
  Physics\/} {\bf 118} 173105

\bibitem{cherevko2009gold}
Cherevko S and Chung C~H 2009 {\em Sensors and Actuators B: Chemical\/} {\bf
  142} 216--23

\bibitem{kabashin2009plasmonic}
Kabashin A, Evans P, Pastkovsky S, Hendren W, Wurtz G, Atkinson R, Pollard R,
  Podolskiy V and Zayats A 2009 {\em Nature materials\/} {\bf 8} 867

\bibitem{caldwell2011plasmonic}
Caldwell J~D, Glembocki O, Bezares F~J, Bassim N~D, Rendell R~W, Feygelson M,
  Ukaegbu M, Kasica R, Shirey L and Hosten C 2011 {\em ACS nano\/} {\bf 5}
  4046--55

\bibitem{wang2012plasmonic}
Wang K and Crozier K~B 2012 {\em ChemPhysChem\/} {\bf 13} 2639--48

\bibitem{bruggemann2011nanostructured}
Br{\"u}ggemann D, Wolfrum B, Maybeck V, Mourzina Y, Jansen M and
  Offenh{\"a}usser A 2011 {\em Nanotechnology\/} {\bf 22} 265104

\bibitem{navitski2013field}
Navitski A, Lagotzky S, Reschke D, Singer X and M{\"u}ller G 2013 {\em Physical
  Review Special Topics-Accelerators and Beams\/} {\bf 16} 112001

\bibitem{nagaoka2001field}
Nagaoka K, Fujii H, Matsuda K, Komaki M, Murata Y, Oshima C and Sakurai T 2001
  {\em Applied surface science\/} {\bf 182} 12--9

\bibitem{engelberg2018stochastic}
Engelberg E~Z, Ashkenazy Y and Assaf M 2018 {\em Physical Review Letters\/}
  {\bf 120} 124801

\bibitem{clic2018compact}
CLIC, CLICdp, Charles T, Giansiracusa P, Lucas T, Rassool R, Volpi M, Balazs C,
  Afanaciev K, Makarenko V, Patapenka A, Zhuk I {\em et~al.\/} 2018
  \href{https://doi.org/10.23731/CYRM-2018-002}{{\em CERN Yellow Reports\/}
  {\bf 2}} (\textit{Preprint} \href{https://arxiv.org/abs/1812.06018}{{\tt
  1812.06018}}) \urlprefix\url{http://dx.doi.org/10.23731/CYRM-2018-002}

\bibitem{timko2014from}
Timko H, Sjobak K, Mether L, Calatroni S, Djurabekova F, Matyash K, Nordlund K,
  Schneider R and Wuensch W 2014 Submitted to Contributions to Plasma Physics

\bibitem{wuensch2013advances}
Wuensch W 2013 \textit{Preprint} CERN-OPEN-2014-028
  \urlprefix\url{http://dx.doi.org/10.1142/9789814602105_0003}

\bibitem{tsong1975direct}
Tsong T and Kellogg G 1975 {\em Physical Review B\/} {\bf 12} 1343

\bibitem{wang1982field}
Wang S and Tsong T 1982 {\em Physical Review B\/} {\bf 26} 6470

\bibitem{hakkinen1993roughening}
H{\"a}kkinen H, Merikoski J, Manninen M, Timonen J and Kaski K 1993 {\em
  Physical Review Letters\/} {\bf 70} 2451

\bibitem{wang2001kinetic}
Wang L and Clancy P 2001 {\em Surface Science\/} {\bf 473} 25--38

\bibitem{lam2002competing}
Lam C~H, Lee C~K and Sander L~M 2002 {\em Physical Review Letters\/} {\bf 89}
  216102

\bibitem{zhang2004kinetic}
Zhang P, Zheng X, Wu S, Liu J and He D 2004 {\em Vacuum\/} {\bf 72} 405--10

\bibitem{kara2009off}
Kara A, Trushin O, Yildirim H and Rahman T~S 2009 {\em Journal of Physics:
  Condensed Matter\/} {\bf 21} 084213

\bibitem{nandipati2012off}
Nandipati G, Kara A, Shah S~I and Rahman T~S 2012 {\em Journal of Computational
  Physics\/} {\bf 231} 3548--60

\bibitem{jansson2016long}
Jansson V, Baibuz E and Djurabekova F 2016
  \href{https://doi.org/10.1088/0957-4484/27/26/265708}{{\em Nanotechnology\/}
  {\bf 27} 265708} (\textit{Preprint}
  \href{https://arxiv.org/abs/1508.06870}{{\tt 1508.06870}})
  \urlprefix\url{https://doi.org/10.1088/0957-4484/27/26/265708}

\bibitem{kimocs}
 2014 {Kimocs --- a Kinetic Monte Carlo simulation code for surfaces}
  {Available under the terms of the GNU General Public License.}
  \urlprefix\url{https://gitlab.com/vjansson/Kimocs}

\bibitem{vigonski2018au}
Vigonski S, Jansson V, Vlassov S, Polyakov B, Baibuz E, Oras S, Aabloo A,
  Djurabekova F and Zadin V 2018
  \href{https://doi.org/https://doi.org/10.1088/1361-6528/aa9a1b}{{\em
  Nanotechnology\/} {\bf 29} 015704} (\textit{Preprint}
  \href{https://arxiv.org/abs/1709.09104}{{\tt 1709.09104}})
  \urlprefix\url{10.1088/1361-6528/aa9a1b}

\bibitem{zhao2016formation}
Zhao J, Baibuz E, Vernieres J, Grammatikopoulos P, Jansson V, Nagel M,
  Steinhauer S, Sowwan M, Kuronen A, Nordlund K {\em et~al.\/} 2016
  \href{https://doi.org/10.1021/acsnano.6b01024}{{\em ACS nano\/} {\bf 10}
  4684--94} \urlprefix\url{https://doi.org/10.1021/acsnano.6b01024}

\bibitem{vurpillot2018simulation}
Vurpillot F, Parviainen S, Djurabekova F, Zanuttini D and Gervais B 2018 {\em
  Materials Characterization\/} {\bf 146} 336--46

\bibitem{djurabekova2011atomistic}
Djurabekova F, Parviainen S, Pohjonen A and Nordlund K 2011 {\em Physical
  Review E\/} {\bf 83} 026704

\bibitem{parviainen2011molecular}
Parviainen S, Djurabekova F, Pohjonen A and Nordlund K 2011 {\em Nuclear
  Instruments and Methods in Physics Research Section B: Beam Interactions with
  Materials and Atoms\/} {\bf 269} 1748--51

\bibitem{veske2016electrodynamics}
Veske M, Parviainen S, Zadin V, Aabloo A and Djurabekova F 2016
  \href{https://doi.org/10.1088/0022-3727/49/21/215301}{{\em Journal of Physics
  D: Applied Physics\/} {\bf 49} 215301} (\textit{Preprint}
  \href{https://arxiv.org/abs/1601.00407}{{\tt 1601.00407}})
  \urlprefix\url{https://doi.org/10.1088/0022-3727/49/21/215301}

\bibitem{wang2004novel}
Wang P~I, Zhao Y, Wang G and Lu T 2004 {\em Nanotechnology\/} {\bf 15} 218

\bibitem{amram2015capillary}
Amram D, Kovalenko O, Klinger L and Rabkin E 2015 {\em Scripta Materialia\/}
  {\bf 109} 44--7

\bibitem{soisson2007cu}
Soisson F and Fu C~C 2007 {\em Physical Review B\/} {\bf 76} 214102

\bibitem{vincent2008precipitation}
Vincent E, Becquart C, Pareige C, Pareige P and Domain C 2008 {\em Journal of
  Nuclear Materials\/} {\bf 373} 387--401

\bibitem{castin2011modeling}
Castin N, Pascuet M~I and Malerba L 2011 {\em The Journal of chemical
  physics\/} {\bf 135} 064502

\bibitem{baibuz2018migration}
Baibuz E, Vigonski S, Lahtinen J, Zhao J, Jansson V, Zadin V and Djurabekova F
  2018 \href{https://doi.org/10.1016/j.commatsci.2017.12.054}{{\em
  Computational Materials Science\/} {\bf 146}(C) 287--302}
  \urlprefix\url{https://doi.org/10.1016/j.commatsci.2017.12.054}

\bibitem{jansson2020tungsten}
Jansson V, Kyritsakis A, Vigonski S, Baibuz E, Zadin V, Aabloo A and
  Djurabekova F 2020 \href{https://doi.org/10.1088/1361-651X/ab7151}{{\em
  Modelling and Simulation in Materials Science and Engineering\/} {\bf 28}
  035011} (\textit{Preprint} \href{https://arxiv.org/abs/1909.03519}{{\tt
  1909.03519}}) \urlprefix\url{https://doi.org/10.1088/1361-651X/ab7151}

\bibitem{olewicz2014coexistence}
Olewicz T, Antczak G, Jurczyszyn L, Lyding J~W and Ehrlich G 2014
  \href{https://doi.org/10.1103/PhysRevB.89.235408}{{\em Phys. Rev. B\/} {\bf
  89}(23) 235408}

\bibitem{jackson1975classical}
Jackson J 1975 {\em Classical electrodynamics\/} (Wiley, New York) ISBN
  9780471431329

\bibitem{press1992numerical}
Press W~H, Teukolsky S~A, Vetterling W~T and Flannery B~P 1992 {\em Numerical
  recipes in C: Art of Scientific Computing, 2nd ed.\/} (Cambridge University
  Press, New York)

\bibitem{fujita2007mechanism}
Fujita S and Shimoyama H 2007 {\em Physical Review B\/} {\bf 75} 235431

\bibitem{ovito}
Stukowski A 2010 {\em Modelling and Simulation in Materials Science and
  Engineering\/} {\bf 18} 015012 \urlprefix\url{http://ovito.org/}

\bibitem{lahtinen2018artificial}
Kimari J, Jansson V, Vigonski S, Baibuz E, Domingos R, Zadin V and Djurabekova
  F 2020 {\em {arXiv:1806.02976 [physics.comp-ph]}\/}
  \urlprefix\url{https://arxiv.org/abs/1806.02976}

\end{thebibliography}

\end{document}